%% file: main.tex
\documentclass[fleqn,usenatbib,useAMS]{mnras}

\usepackage{mathptmx}

\usepackage[T1]{fontenc}
\usepackage{ae,aecompl}

\usepackage{graphicx}	
\usepackage{amsmath}	
\usepackage{amssymb}	
\usepackage{bm}		
\usepackage{pdflscape}	
\usepackage{color}
\usepackage{xcolor}
\usepackage{exscale}
\usepackage{natbib}
\usepackage{rotating}
\usepackage{rotate}
\usepackage{afterpage}
\usepackage{booktabs,threeparttable}
\usepackage{relsize}
\usepackage{lscape}
\usepackage{tabularx}
\usepackage{longtable}
\usepackage{ltxtable}
\usepackage{txfonts}
\usepackage{bbm}
\usepackage{widetext}
\usepackage{subfig}
\usepackage{multirow}
\usepackage{xurl}
\usepackage{bm}

\usepackage{soul}
\input{journaldef.tex}

\newcommand{\ie}{i.e.}
\newcommand{\eg}{e.g.}

\newcommand{\ee}{\mathrm{e}}
\newcommand{\bv}[1]{\mathbfit{#1}}
\newcommand{\bmat}[1]{\mathbfss{#1}}
\newcommand{\x}{$\times$}
\newcommand{\btop}{{\bm{\top}}}
\newcommand{\bdot}{{\bm{\cdot}}}
\newcommand{\cov}{{\rm Cov}}

\title[LSST+SO Simulated Likelihood Analysis]{Cosmology from Clustering, Cosmic Shear, CMB Lensing, and Cross Correlations: Combining Rubin Observatory and Simons Observatory}

\author[X. Fang et al.]{Xiao Fang,$^{1,3}$\thanks{E-mail: xfang@berkeley.edu} Tim Eifler$^{1}$, Emmanuel Schaan$^{2,3}$, Hung-Jin Huang$^{1}$, Elisabeth Krause$^{1,4}$, Simone Ferraro$^{2,3}$
\\
$^{1}$Department of Astronomy and Steward Observatory, University of Arizona, 933 North Cherry Avenue, Tucson, AZ 85721, USA\\
$^{2}$Lawrence Berkeley National Laboratory, One Cyclotron Road, Berkeley, CA 94720, USA\\
$^{3}$Berkeley Center for Cosmological Physics, UC Berkeley, CA 94720, USA\\
$^{4}$Department of Physics, University of Arizona, 1118 E. Fourth Street, Tucson, AZ 85721, USA
}

\begin{document}

\date{Accepted . Received ; in original form }

\pagerange{\pageref{firstpage}--\pageref{lastpage}} \pubyear{2021}

\maketitle

\label{firstpage}

\begin{abstract} 
In the near future, the overlap of the Rubin Observatory Legacy Survey of Space and Time (LSST) and the Simons Observatory (SO) will present an ideal opportunity for joint cosmological dataset analyses. In this paper we simulate the joint likelihood analysis of these two experiments using six two-point functions derived from galaxy position, galaxy shear, and CMB lensing convergence fields. Our analysis focuses on realistic noise and systematics models and we find that the dark energy Figure-of-Merit (FoM) increases by 53\% (92\%) from LSST-only to LSST+SO in Year 1 (Year 6). We also investigate the benefits of using the same galaxy sample for both clustering and lensing analyses, and find the choice improves the overall signal-to-noise by $\sim30-40\%$, which significantly improves the photo-z calibration and mildly improves the cosmological constraints. Finally, we explore the effects of catastrophic photo-z outliers finding that they cause significant parameter biases when ignored. We develop a new mitigation approach termed ``island model'', which corrects a large fraction of the biases with only a few parameters while preserving the constraining power.

\end{abstract}


\renewcommand{\thefootnote}{\arabic{footnote}}
\setcounter{footnote}{0}

\section{Introduction}
\label{sec:intro}
The large-scale structure in the Universe has been a major source of information about the structure growth and the cosmic expansion history, which allows us to test theories of gravity, the mass and number of species of neutrinos, and the nature of dark energy and dark matter. These science questions have motivated the development of a series of ongoing and upcoming galaxy survey experiments, including the Kilo-Degree Survey \citep[KiDS,][]{2017MNRAS.465.1454H,2021A&A...646A.140H}, the Dark Energy Survey \citep[DES,][]{2018PhRvD..98d3526A,2021arXiv210513549D}, and the Hyper Suprime-Cam \citep[HSC,][]{2019PASJ...71...43H}, the Dark Energy Spectroscopic Instrument \citep[DESI,][]{2016arXiv161100036D}, the Vera C. Rubin Observatory Legacy Survey of Space and Time \citep[LSST,][]{2019ApJ...873..111I}, the Nancy Grace Roman Space Telescope \citep{2019arXiv190205569A}, the Euclid \citep{Euclid_WhitePaper}, and the Spectro-Photometer for the History of the Universe,
Epoch of Reionization, and Ices Explorer \citep[SPHEREx,][]{2014arXiv1412.4872D}. The results from KiDS, DES, and HSC have demonstrated the approach of increasing the overall constraining power by jointly analysing different cosmological probes, notably the galaxy clustering and weak lensing statistics. Such a ``multi-probe analysis'' approach will continue to play a key role in extracting and combining cosmological information from future galaxy survey experiments.

The cosmic microwave background (CMB) provides a screenshot of the Universe at its infancy, which not only carries a wealth of information about the energy components of the Universe, but also acts as an anchor for the structure growth at redshift $z$ as high as $\sim 1100$. Within the $\Lambda$CDM model, the difference in $S_8$ parameter between high-$z$ CMB measurements from Planck \citep{2020A&A...641A...6P} and many low-$z$ measurements, such as KiDS \citep{2018MNRAS.474.4894J,2021A&A...646A.140H}, DES \citep{2018PhRvD..98d3526A,2019PhRvD.100b3541A,2021arXiv210513549D}, and unWISE and Planck CMB lensing tomography \citep{2021arXiv210503421K}, has been heavily studied but still remains uncertain. A series of ongoing and upcoming CMB experiments with increasing sensitivities, including the Atacama Cosmology Telescope \citep[ACT,][]{2020JCAP...12..047A,2020JCAP...12..045C}, the South Pole Telescope \citep[SPT,][]{2014SPIE.9153E..1PB,2021arXiv210101684D}, the Simons Observatory \citep[SO,][]{2018SPIE10708E..04G,2019JCAP...02..056A}, and the CMB Stage-4 \citep[S4,][]{2016arXiv161002743A,2019arXiv190704473A}, together with various wider and deeper galaxy surveys, will undoubtedly increase the significance of the tension/agreement and potentially reveal deviations from the standard $\Lambda$CDM model in the near future.

The CMB lensing power spectrum has become an increasingly powerful cosmological probe thanks to sensitivity improvements of CMB experiments in temperature and polarisation. The recent constraints on the matter density parameter $\Omega_m$ and the amplitude of the (linear) power spectrum on the scale of $8h^{-1}$Mpc, $\sigma_8$, from the Planck CMB lensing alone have achieved precision comparable to DES Y1 weak lensing results \citep{2020A&A...641A...6P}. Cross correlations between CMB lensing convergence and galaxy positions or galaxy shapes have been measured with increasing signal-to-noise ratios \citep[\eg,][]{2016MNRAS.456.3213G,2017MNRAS.464.2120S,2019PhRvD.100d3501O,2019PhRvD.100d3517O,2019PhRvD.100b3541A,2021arXiv210503421K,2021arXiv210315862M,2021A&A...649A.146R,2021MNRAS.501.1481H,2021MNRAS.501.6181K}. The CMB lensing signal carries information about the growth history from redshift $\sim 1-3$, valuable to the determination of dark energy properties and highly complementary to galaxy weak lensing and galaxy clustering. Free from intrinsic alignment contamination and with an independent set of systematic errors, CMB lensing has also been shown to improve the calibration of various systematics in galaxy surveys when combined with so-called ``3\x2pt probes'', which include projected galaxy clustering, galaxy-galaxy lensing, and cosmic shear \citep[\eg,][]{2017PhRvD..95l3512S,2020JCAP...12..001S}. The significant role of such cross-correlations in the context of CMB-S4 is emphasised in \cite{2016arXiv161002743A}.

We focus on the synergies between LSST and SO, specifically we explore the cosmological information gains when combining SO's CMB lensing convergence with LSST's 3\x2pt probes. We refer the combination of the CMB lensing and 3\x2pt probes as the 6\x2pt, since 3 more types of two-point correlations are included. Note that other 6\x2pt combinations are possible, such as replacing CMB lensing with cluster density \citep{2021PhRvL.126n1301T}. We summarise the key characteristics of the LSST and SO experiments below.

\paragraph*{Vera C. Rubin Observatory's Legacy Survey of Space and Time (LSST)} aims to start commissioning and science verification in 2022 and with the goal to be fully operational in 2023. Equipped with a 6.5 m (effective diameter) primary mirror, a 9.6 deg$^2$ field of view, and a  3.2-gigapixel camera, LSST's observing strategy \citep{2019ApJ...873..111I, 2018arXiv181200515L} is to rapidly and repeatedly cover its footprint ($\sim$ 18,000 deg$^2$) in 6 optical bands (320 nm-1050 nm). With a single exposure depth of 24.7 r-band magnitude (5$\sigma$ point source), the 10 years of operations will achieve an overall depth of 27.5 r-band magnitude. 

The LSST-Dark Energy Science Collaboration (DESC) is tasked to conduct the dark energy data analysis based on LSST data. The performance of an LSST-DESC analysis given specific analysis choices is explored in the DESC-Science Requirements Document \citep[DESC-SRD,][]{2018arXiv180901669T}, and the analysis choices of that document have been adopted for this paper as well. LSST will be a highly synergistic dataset for all surveys of the coming decade, in particular with the NASA Roman Space Telescope \citep[\eg,][]{2021MNRAS.tmp..589E} and the ESA/NASA Euclid satellite mission \citep[\eg,][]{capak2019enhancing}.

\paragraph*{Simons Observatory}
The Simons Observatory (SO, \citealp{Galitzki18, Ade19}) is a CMB experiment under construction in the Atacama desert in Chile, at an altitude of 5,200~m.
It is designed to observe the microwave sky in six frequency bands centred around 30, 40, 90, 150, 230, and 290~GHz, in order to separate the CMB from Galactic and extragalactic foregrounds.

The observatory will include one 6~m large-aperture telescope (LAT, \citealp{Xu21, Parshley18}) and three small-aperture 0.5~m telescopes (SATs, \citealp{Ali20}).
The LAT will produce temperature and polarisation maps of the CMB with $\sim$arcminute resolution over 40\% of the sky, with a sensitivity of $\sim$6 $\mu$K$\cdot$arcmin when combining 90 and 150~GHz bands.
These wide deep maps will be the key input to measure CMB lensing with SO.
\newline

In this paper, we compare constraints on cosmological and systematic parameters from the LSST+SO 6\x2pt probes and from LSST-only 3\x2pt probes. Since both experiments start approximately at the same time, we study how the constraining power increases from Y1 to Y6 as the experiments increase depth and overlapping survey area. As the most constraining joint analysis we consider LSST's Y6 data and SO's final data product (SO Y5). 

We begin the paper with describing our analysis choices and outlining the theoretical modelling details in Section~\ref{sec:analysis}. Section~\ref{sec:forecast} contains the results of our simulated likelihood analyses, where we consider two cases as the lens sample: First, we explore the same lens sample as defined in the DESC-SRD and second we consider the source galaxy sample acting as the lens sample, similar to \cite{2020JCAP...12..001S}. Interestingly, their recent Fisher analysis has found large improvements in photometric self-calibrations and mild improvements in cosmological parameter constraints using this choice. We examine the same idea by running a simulated MCMC likelihood analysis using Non-Gaussian covariances and a more complete systematics model, e.g., we include baryonic physics effects and intrinsic alignment and later study catastrophic outliers based on realistic photo-z simulations. 

In Section~\ref{sec:outlier}, we study the impact of catastrophic photo-z outliers on LSST's cosmological parameter inference. We use the simulated photo-z catalogues from \cite{2018AJ....155....1G, 2020AJ....159..258G} as our realistic description of LSST photo-z and quantify cosmological biases when ignoring these systematic effects. We further develop a mitigation strategy, termed the ``island model'', which is based on the idea to identify the most relevant features, aka ``islands'', in the photo-z vs true redshift diagram and then marginalise over the amplitude in these features. We conclude in Section~\ref{sec:conclu}.

\section{Joint-Survey Multi-Probe Analysis}\label{sec:analysis}
Our joint-survey multi-probe forecast assumes a Gaussian likelihood of the data $\bv{D}$ given a point $\bv{p}$ in cosmological and nuisance parameter space,
\begin{equation}
    L(\bv{D}|\bv{p}) \propto \ee^{-\chi^2/2}~,~~{\rm where}~~\chi^2=[\bv{D}-\bv{M}(\bv{p})]^\btop\bmat{C}^{-1}[\bv{D}-\bv{M}(\bv{p})]~,
\end{equation}
$\bv{M}$ is the model vector and $\bmat{C}$ is the covariance matrix. In this paper, the data vector is the concatenation of 6 two-point functions, specifically, the angular power spectra measured from the observed galaxy density and lensing convergence fields by LSST and the reconstructed CMB lensing convergence field by SO. The detailed analysis choices are described in Section~\ref{ssec:ana-choices} and the modelling of these two-point functions is detailed in Section~\ref{ssec:2pt}. In Section~\ref{ssec:sys}, we describe the modelling of systematic effects included in this analysis. Finally, we model the corresponding covariance matrix analytically in Section~\ref{ssec:cov}.

Throughout our analyses, we use \textsc{CosmoLike}\footnote{https://github.com/CosmoLike} \citep{2014MNRAS.440.1379E,2017MNRAS.470.2100K} for the modelling and inference process.  Data vectors \bv{D} are computed at the fiducial parameter values (Table \ref{tab:params}) and in our fiducial cosmology (standard $\Lambda$CDM with massless neutrinos). The code setup is comparable to simulated analyses run in the context of Roman Space Telescope to explore multi-probe strategies \citep{2021MNRAS.tmp.1608E,2021MNRAS.tmp..589E} and in the DESC-SRD for the static probes \citep{2018arXiv180901669T}. We use an extended version of \textsc{CosmoCov}\footnote{https://github.com/CosmoLike/CosmoCov} \citep{2020MNRAS.497.2699F} for the analytic covariances.

\subsection{Analysis Choices}\label{ssec:ana-choices}
\begin{figure}
    \centering
    \includegraphics[width=\columnwidth]{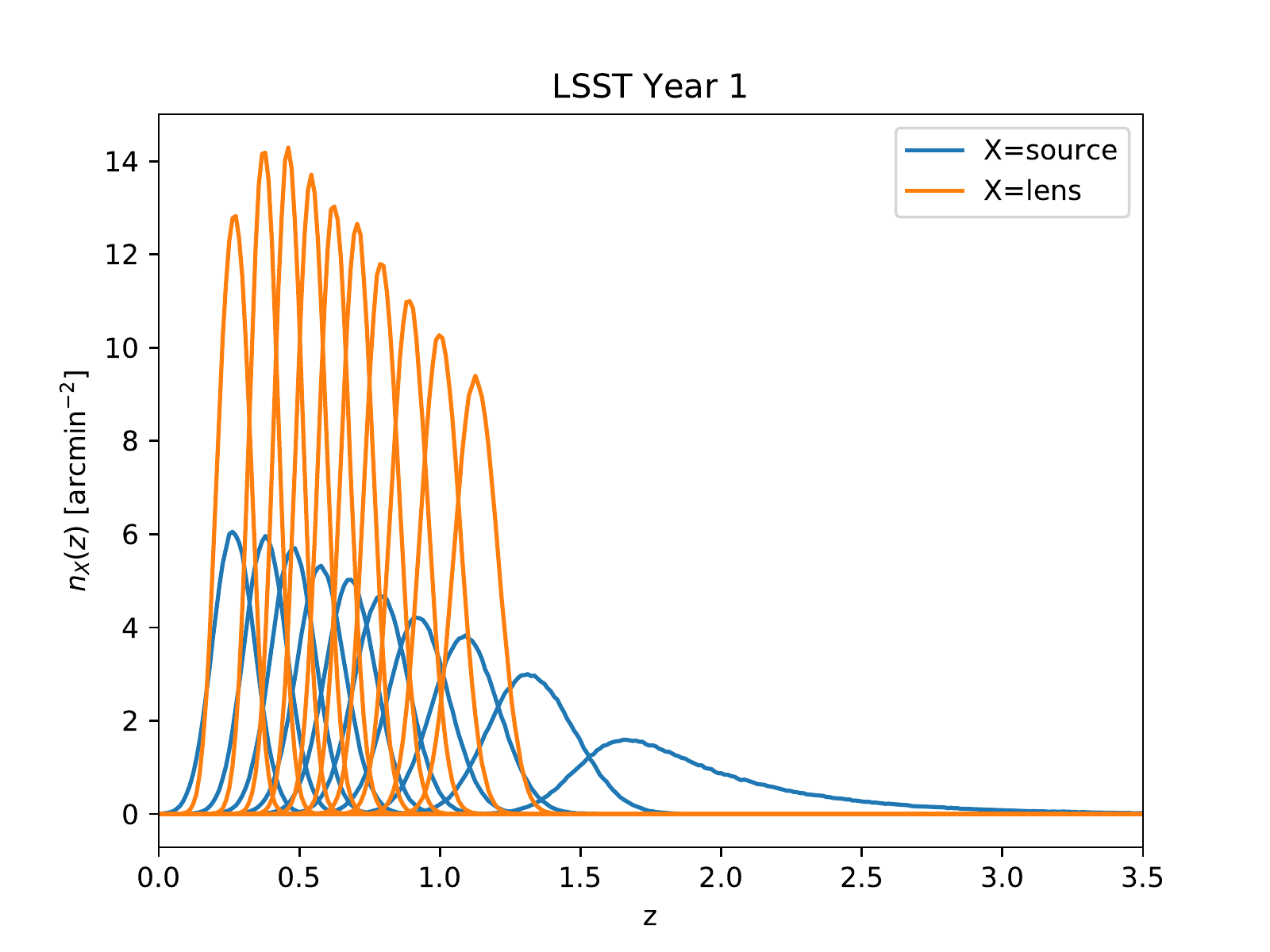}
    \includegraphics[width=\columnwidth]{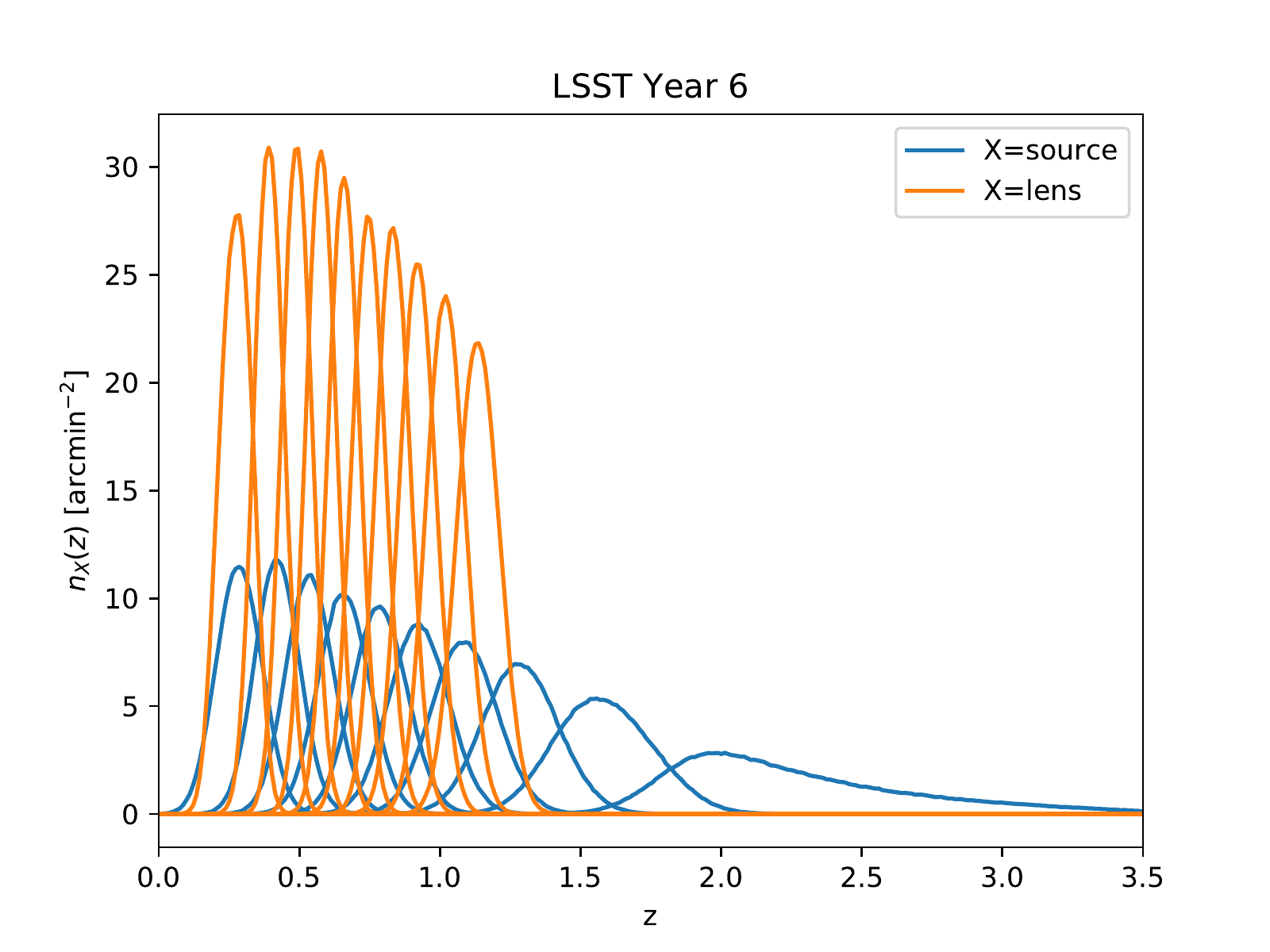}
    \caption{The equally binned true redshift distributions $n_X(z)=\frac{dN_X}{dz d\Omega}$ of the source and lens samples of LSST Year 1 (upper panel) and Year 6 (lower panel), normalised by the corresponding effective number density $\bar{n}_X^i$. A fiducial Gaussian photo-z error has been convolved with each tomographic bin as described by Eq.~(\ref{eq:binned-nz-true}).}
    \label{fig:nzs}
\end{figure}

We follow the LSST-DESC SRD (\citealp{2018arXiv180901669T}) in choosing the galaxy samples for the galaxy clustering and weak lensing analyses. The LSST Y1 survey will cover a survey area of 12,300 deg$^2$, with an i-band depth $i_{\rm depth}=25.1\,$mag for the weak lensing analysis (\ie, the source sample) and and i-band limiting magnitude $i_{\rm lim}=i_{\rm depth}-1=24.1\,$mag for the large-scale structure (LSS) or the clustering analysis (\ie, the lens sample). The SO Y1 is assumed to observe 40\% of the sky (16,500 deg$^2$), hence we restrict our analysis to the overlapping footprint which in this case is LSST's survey area. We further assume that LSST Y6 will cover the final SO Y5 area (again 40\% of the sky, but deeper than Y1). We also assume an i-band depth $i_{\rm depth}=26.1\,$mag for the weak lensing analysis, and i-band limiting magnitude $i_{\rm lim}=i_{\rm depth}-1=25.1\,$mag for the clustering analysis.

For both the lens and source samples, we parameterise our photometric redshift distribution as
\begin{equation}
    n_X(z_{\rm ph}) \equiv \frac{dN_X}{dz_{\rm ph}d\Omega} \propto z_{\rm ph}^2\exp[-(z_{\rm ph}/z_0)^\alpha]~,~~X\in\{{\rm lens,~source}\}
    \label{eq:nz-true}
\end{equation}
normalised by the effective number density $\bar{n}_X$. $N_X$ is the number counts of lens/source galaxies, $z_{\rm ph}$ is the photometric redshift, $\Omega$ is the solid angle. The parameter values of $\bar{n}_X,~z_0,~\alpha$ are taken from the DESC-SRD and are given in Table~\ref{tab:nz}. Other values in Table~\ref{tab:nz} that are slightly different from the DESC-SRD were computed for the more recent optimisation studies of LSST observing strategies \citep{2018arXiv181200515L} and were updated on the LSST-DESC's Observing Strategy GitHub page\footnote{\url{https://github.com/LSSTDESC/ObsStrat/tree/static/static}}. We impose a high-$z$ cut $z_{\rm max}=3.5$ for source galaxies and $z_{\rm max}=1.2$ for lens galaxies following the DESC-SRD. We further divide each galaxy sample into 10 equally populated tomographic bins, \ie, the effective number density of each bin $\bar{n}^i_X=\bar{n}_X/10$, as shown Figure~\ref{fig:nzs}.

\begin{table}
    \centering
    \begin{tabular}{|c|c|c|c|}
    \hline
         & Parameter  & LSST Y1 & LSST Y6 \\
         \hline
    lens & $\bar{n}_{\rm lens}$  & 18.0  & 41.1 \\
         & $z_0$ & 0.260 & 0.274 \\
         & $\alpha$ & 0.942 & 0.907 \\
         \hline
    source & $\bar{n}_{\rm source}$ & 11.2  & 23.2 \\
         & $z_0$ & 0.191 & 0.178 \\
         & $\alpha$ & 0.870 & 0.798 \\
         \hline
    \end{tabular}
    \caption{Parameters for the lens and source galaxy samples used in LSST Y1 and Y6.}
    \label{tab:nz}
\end{table}
\begin{table}
\footnotesize
    \centering
    \begin{tabular}{|l c c c|}
    \hline
    Parameters & Fiducial & \multicolumn{2}{c|}{Prior} \\
    \hline
    \multicolumn{4}{|l|}{\textbf{Survey}} \\
    $\Omega_{\rm s}$ (deg$^2$) & Y1: 12300 & \multicolumn{2}{c|}{fixed} \\
    & Y6: 16500 & \multicolumn{2}{c|}{fixed} \\
    $\sigma_e$ & 0.26/component & \multicolumn{2}{c|}{fixed} \\
    \hline
    \multicolumn{2}{|l}{\textbf{Cosmology}} & \multicolumn{2}{c|}{flat} \\
    $\Omega_{\rm m}$ & 0.3156 & \multicolumn{2}{c|}{[0.05, 0.6]}\\
    $\sigma_8$ & 0.831 & \multicolumn{2}{c|}{[0.5, 1.1]} \\
    $n_{\rm s}$ & 0.9645 & \multicolumn{2}{c|}{[0.85, 1.05]}\\
    $\Omega_b$ & 0.0492 & \multicolumn{2}{c|}{[0.04, 0.055]}\\
    $h_0$ & 0.6727 & \multicolumn{2}{c|}{[0.4, 0.9]} \\
    $w_0$ & -1 & \multicolumn{2}{c|}{[-2, 0]} \\
    $w_a$ & 0 & \multicolumn{2}{c|}{[-2.5, 2.5]} \\
    $\mu_0$ & 0 & \multicolumn{2}{c|}{[-3, 3] $^a$} \\
    $\Sigma_0$ & 0 & \multicolumn{2}{c|}{[-3, 3] $^a$} \\
    \hline
    \multicolumn{2}{|l}{\textbf{Galaxy Bias}} & \multicolumn{2}{c|}{flat}\\
    $b^i$ & $0.95/G(\langle z^i\rangle)$ & \multicolumn{2}{c|}{[0.4, 3] ([0.4, 5] for ``lens=source'')} \\
    \hline
    \multicolumn{2}{|l}{\textbf{Photo-$z$}} & Y1 & Y6 \\
    $\Delta_{z,\rm lens}^i$ & 0 & $\mathcal{N}(0, 0.002^2)$ & $\mathcal{N}(0, 0.001^2)$ \\
    $\sigma_{z,\rm lens}$ & 0.03 & $\mathcal{N}(0.03, 0.006^2)$ & $\mathcal{N}(0.03, 0.003^2)$ \\
    $\Delta_{z,\rm source}^i$ & 0 & $\mathcal{N}(0, 0.002^2)$ & $\mathcal{N}(0, 0.001^2)$ \\
    $\sigma_{z,\rm source}$ & 0.05 & $\mathcal{N}(0.05, 0.006^2)$ & $\mathcal{N}(0.05, 0.003^2)$ \\
    \hline
    \multicolumn{2}{|l}{\textbf{Shear Calibration}} & Y1 & Y6 \\
    $m^i$ & 0 & $\mathcal{N}(0, 0.013^2)$ & $\mathcal{N}(0, 0.003^2)$ \\
    \hline
    \multicolumn{2}{|l}{\textbf{IA}} & \multicolumn{2}{c|}{Gaussian} \\
    $A_{\rm IA}$ & 5.95 & \multicolumn{2}{c|}{$\mathcal{N}(5.95, 3.0^2)$ within [0, 10]}\\
    $\beta_{\rm IA}$ & 1.1 & \multicolumn{2}{c|}{$\mathcal{N}(1.1, 1.2^2)$ within [-4, 6]}\\
    $\eta_{\rm IA}$ & -0.47 & \multicolumn{2}{c|}{$\mathcal{N}(-0.47, 3.8^2)$ within [-10, 10]}\\
    $\eta_{\rm IA}^{\rm high-z}$ & 0 & \multicolumn{2}{c|}{$\mathcal{N}(0, 2.0^2)$ within [-1, 1]} \\
    \hline
    \multicolumn{2}{|l}{\textbf{Baryon}} & Y1 & Y6 \\
    $Q_1$ & 0 & $\mathcal{N}(0, 16^2)$ & $\mathcal{N}(0, 29^2)$\\
    $Q_2$ & 0 & $\mathcal{N}(0, 1.9^2)$ & $\mathcal{N}(0, 3.5^2)$\\
    $Q_3$ & 0 & $\mathcal{N}(0, 0.7^2)$ & $\mathcal{N}(0, 1.7^2)$\\
    \hline
    \end{tabular}
    \caption{A list of the parameters characterising the surveys, cosmology and systematics. ``Y1'' stands for LSST Y1 + SO Y1 and ``Y6'' for LSST Y6 + SO Y5. The fiducial values are used for generating the simulated data vectors, and the priors are used in the sampling. Flat priors are described by [minimum, maximum], and Gaussian priors are described through the normal distribution $\mathcal{N}(\mu, \sigma^2)$. For the ``lens=source'' sample choice discussed in Section~\ref{sec:1sample}, $\Delta_{z,{\rm lens}}^i=\Delta_{z,{\rm source}}^i$, $\sigma_{z,{\rm lens}}=\sigma_{z,{\rm source}}$, and the re-derived Gaussian priors  of $Q_i$'s are approximately the same as their DESC-SRD version.\\
    $^a$ $(\mu_0,\Sigma_0)$ are only varied when we compute the constraints for MG models, while kept fixed at fiducial values in the other analyses.}
    \label{tab:params}
\end{table}

\subsection{Two-point function modelling details}\label{ssec:2pt}
We consider LSST cosmic shear, galaxy-galaxy lensing and photometric galaxy clustering probes, which in combination form a so-called 3\x2pt analysis. Adding the SO CMB lensing convergence field, we can extend the data vector with 3 more two-point functions: galaxy density-CMB lensing, galaxy shape-CMB lensing, and the CMB lensing auto correlation. In this subsection, we summarise the computation of angular (cross) power spectra for the different probes. We use capital Roman subscripts to denote observables, $A,B\in\lbrace\delta_{\rm g}, \kappa_{\rm g},\kappa_{\rm CMB}\rbrace$, where $\delta_{\rm g}$ denotes the density contrast of lens galaxies, $\kappa_{\rm g}$ the lensing convergence of source galaxies, and $\kappa_{\rm CMB}$ the CMB lensing convergence.

\subsubsection{Modelling of Probes}\label{sssec:probe-modeling}
Within the Limber approximation (\citealp{1953ApJ...117..134L,2008PhRvD..78l3506L}; see \citealp{2020JCAP...05..010F} for potential impact in current and near future surveys), the angular power spectrum between redshift bin $i$ of observable $A$ and redshift bin $j$ of observable $B$ at Fourier mode $\ell$ is given by
\begin{equation}
C_{AB}^{ij}(\ell)=\int d\chi\frac{q_A^i(\chi)q_B^j(\chi)}{\chi^2}P_{AB}\left(\frac{\ell+1/2}{\chi},z(\chi)\right)~,
\end{equation}
where $\chi$ is the comoving distance, $P_{AB}(k,z)$ is the 3D probe-specific power spectra, $q_{A}^i(\chi),q_{B}^j(\chi)$ are weight functions of the observables $A,B$ given by
\begin{align}
    &q_{\delta_{\rm g}}^i(\chi) = \frac{n_{\rm lens}^i(z(\chi))}{\bar{n}^i_{\rm lens}}\frac{dz}{d\chi}~,\\
    &q_{\kappa_{\rm g}}^i(\chi) = \frac{3H_0^2\Omega_m}{2c^2}\frac{\chi}{a(\chi)}\int_{\chi_{\rm min}^i}^{\chi_{\rm max}^i}\,d\chi'\frac{n_{\rm source}^i(z(\chi'))}{\bar{n}_{\rm source}^i}\frac{dz}{d\chi'}\frac{\chi'-\chi}{\chi'}~, \\
    &q_{\kappa_{\rm CMB}}(\chi) = \frac{3H_0^2\Omega_m}{2c^2}\frac{\chi}{a(\chi)}\frac{\chi^*-\chi}{\chi^*}~,
\end{align}
where $\chi_{\rm min/max}^i$ are the minimum and maximum comoving distance of the redshift bin $i$, $a(\chi)$ is the scale factor, $\Omega_m$ the matter density fraction at present, $H_0$ the Hubble constant, $c$ the speed of light, and $\chi^*$ the comoving distance to the surface of last scattering. Note that the weight function of $\kappa_{\rm CMB}$ does not depend on redshift bins.

The 3D probe-specific power spectra $P_{AB}(k,z)$ are related to the nonlinear matter power spectrum $P_{\delta\delta}(k,z)$, where $\delta$ is the nonlinear matter density contrast. For different cases, $P_{AB}$ can be determined as
\begin{align}
    &P_{\delta_{\rm g}B}(k,z) = b_{\rm g}(z)P_{\delta B}(k,z)~,\\
    &P_{\kappa_{\rm g}B}(k,z) = P_{\kappa_{\rm CMB}B}(k,z) = P_{\delta B}(k,z)~,\\
    &P_{AB}(k,z) = P_{BA}(k,z)~,
\end{align}
where we have assumed that the galaxy density contrast is proportional to the nonlinear matter density contrast, weighted by an effective galaxy bias parameter $b_{\rm g}(z)$. This assumption is valid on the large scales considered in our analysis.

We choose 15 logarithmically spaced Fourier mode bins ranging from $\ell_{\rm min}=20$ to $\ell_{\rm max}=3000$ for all two-point functions in the data vector. In order to exclude scales where nonlinear galaxy bias must be modelled, for each lens galaxy redshift bin $i$ we impose an $\ell$-cut $\ell^i_{\rm max}=k_{\rm max}\chi(\langle z^i\rangle)-0.5$ for $\delta_{\rm g}$-$\delta_{\rm g}$, $\delta_{\rm g}$-$\kappa_{\rm g}$, and $\delta_{\rm g}$-$\kappa_{\rm CMB}$ cross spectra, where $k_{\rm max}=0.3 h/$Mpc and $\langle z^i\rangle$ is the mean redshift of the lens bin $i$. This scale choice is consistent with DESC-SRD. More sophisticated modelling would be required in a real analysis to robustly extract cosmological information from smaller scales \citep[\eg,][]{2017JCAP...08..009M,2020PhRvD.102l3522P,2021arXiv210513545P, 2021arXiv210503421K,2021MNRAS.501.1481H,2021MNRAS.501.6181K}. We further exclude $\delta_{\rm g}$-$\kappa_{\rm g}$ combinations where the lens redshift bin is completely behind the source redshift bin.

Our analysis assumes General Relativity (GR) and a flat $w_0$-$w_a$CDM cosmology, \ie, a cold dark matter Universe with a time-varying dark energy component with its equation of state $w(a)$ parameterised as \citep{2001IJMPD..10..213C,2003PhRvL..90i1301L}
\begin{equation}
    w(a) = w_0+w_a(1-a)~.
\end{equation}

Throughout this analysis, we use the \textsc{halofit} fitting formulae \citep{2003MNRAS.341.1311S} revised by \cite{2012ApJ...761..152T} to compute $P_{\delta\delta}(k,z)$, in which we calculate the linear matter power spectrum with the fitting formulae in \cite{1998ApJ...496..605E}.

\subsubsection{Modelling of Modified Gravity (MG)}\label{sssec:mg}
We also consider a MG parameterization with 2 additional parameters \citep[\eg,][]{2008JCAP...04..013A,2010PhRvD..81j3510Z}, following the notation and convention in \cite{2013MNRAS.429.2249S}.

Limited to scalar perturbations and in Newtonian gauge, the perturbed Friedmann–Lemaître–Robertson–Walker (FLRW) metric can be written as
\begin{equation}
    ds^2=[1+2\Psi(\bv{x},t)]dt^2-a^2(t)[1-2\Phi(\bv{x},t)]d\bv{x}^2~,
\end{equation}
where $\Psi$ is the Newtonian potential and $\Phi$ is the spatial curvature potential. In GR, the two scalar potentials $\Psi_{\rm GR},\Phi_{\rm GR}$ may be determined by the matter distribution. In Fourier space, the Poisson equation becomes
\begin{equation}
    k^2\Phi_{\rm GR}(k,a)=-4\pi G a^2\bar{\rho}\delta(k,a)~,
\end{equation}
where $k$ is the wavenumber and $\bar{\rho}$ is the mean density of the Universe. In the absence of anisotropic stress, $\Psi=\Phi$. Thus, the lensing potential $(\Psi+\Phi)$ experienced by relativistic particles is given by
\begin{equation}
    k^2[\Phi_{\rm GR}(k,a)+\Psi_{\rm GR}(k,a)]=-8\pi G a^2\bar{\rho}\delta(k,a)~.
\end{equation}
At present, this is true for GR if we neglect the anisotropic stress generated by the free-streaming of photons and neutrinos.

Generically, MG models introduce non-vanishing anisotropic stress as well as changes to the scalar potentials, which can be characterised by the $(\mu, \Sigma)$ parameters as,
\begin{align}
    \Psi(k,a) &= [1+\mu(a)]\Psi_{\rm GR}(k,a)~,\\
    \Psi(k,a)+\Phi(k,a) &=[1+\Sigma(a)][\Psi_{\rm GR}(k,a)+\Phi_{\rm GR}(k,a)]~,
\end{align}
where $\mu(a)$ and $\Sigma(a)$ are assumed to be scale independent. The $\mu$ parameter modifies the growth of linear density perturbations such that
\begin{equation}
    \delta''(k,a)+\left(\frac{2}{a}+\frac{\ddot{a}}{\dot{a}^2}\right)\delta'(k,a) - \frac{3\Omega_m}{2a^2}[1+\mu(a)]\delta(k,a) = 0~,\\
\end{equation}
which affects $P_{\delta\delta}$ and hence all angular power spectra we consider. The prime $'$ represents the derivative with respect to $a$ and the dot represents the time derivative. The $\Sigma$ parameter affects the lensing potential such that
\begin{equation}
    q_{\kappa_{\rm g}}^i(\chi)=[1+\Sigma(\chi)]q_{\kappa_{\rm g},{\scriptscriptstyle\rm GR}}^i(\chi)~,~~q_{\kappa_{\rm CMB}}(\chi)=[1+\Sigma(\chi)]q_{\kappa_{\rm CMB},{\scriptscriptstyle\rm GR}}(\chi)~.
\end{equation}

We assume the time evolution of $(\mu,\Sigma)$ to scale linearly with the effective dark energy density given by the background dynamics, \ie,
\begin{equation}
    \mu(a)=\mu_0\frac{\Omega_\Lambda(a)}{\Omega_\Lambda}~,~~\Sigma(a)=\Sigma_0\frac{\Omega_\Lambda(a)}{\Omega_\Lambda}~,
\end{equation}
where $\Omega_\Lambda(a)$ is the density parameter for dark energy, $\Omega_\Lambda\equiv\Omega_\Lambda(a=1)$, and $(\mu_0,\Sigma_0)$ represent $(\mu,\Sigma)$ at $a=1$. In GR, $\mu_0=\Sigma_0=0$.

\subsection{Systematics}\label{ssec:sys}
Systematic uncertainties are parameterised through nuisance parameters, whose fiducial values and priors are summarised in Table~\ref{tab:params}. Our simulated analysis includes nuisance parameters similar to those used in the joint Roman Space Telescope - Rubin Observatory forecasts in \cite{2021MNRAS.tmp..589E}, summarised as follows:

\paragraph*{Photometric redshift uncertainties}
We model photometric redshift uncertainties through Gaussian scatter $\sigma_{z,X}$ for lens and source sample each, and a shift parameter $\Delta_{z,X}^i$ for each redshift bin $i$ of the lens and source samples, such that the binned true redshift distribution $n_X^i(z)$ is related to the binned photometric redshift distribution $n_X^i(z_{\rm ph})$ (Eq.~\ref{eq:nz-true}) by
\begin{equation}
    n_X^i(z) = \int_{z_{{\rm min},X}^i}^{z_{{\rm max},X}^i}\frac{dz_{\rm ph}\,n_X^i(z_{\rm ph})}{\sqrt{2\pi}\sigma_{z,X}(1+z_{\rm ph})}\exp\left[-\frac{(z-z_{\rm ph}-\Delta_{z,X}^i)^2}{2[\sigma_{z,X}(1+z_{\rm ph})]^2}\right]~.
\label{eq:binned-nz-true}
\end{equation}
We set the fiducial values of $\Delta_{z,X}^i$ as zero, $\sigma_{z,{\rm lens}}$ as 0.03, and $\sigma_{z,{\rm source}}$ as 0.05, following the DESC-SRD. The resulting distributions are shown in Figure~\ref{fig:nzs}.

We only consider Gaussian photometric redshift uncertainties characterised by shift $\Delta_{z}(z)$ and scatter $\sigma_z(z)$ for the fiducial analysis, but will consider the photo-z outliers in Section~\ref{sec:outlier}. We also assume that the scatter follows a simple redshift scaling $\sigma_{z,X}(1+z)$ and that the shift parameter is one constant $\Delta_{z,X}^i$ per redshift bin. In total, we have 22 photo-z parameters (10 shift parameters and 1 scatter parameter for lens and source sample each). These parameters are marginalised over using Gaussian priors. 

For the source samples, the $\Delta_{z,{\rm source}}^i$ and $\sigma_{z,{\rm source}}$ priors for Y1 are chosen to be consistent with the Y1 requirements in DESC-SRD, while we assume the Y10 requirements for our Y6 analysis. For the lens samples, we choose these priors to be same as the corresponding source sample priors\footnote{In the DESC-SRD version 1, these requirements are specified in the Section 5.1 and 5.2, where the requirements for the source samples are derived for 3\x2pt analysis, but the requirements for lens samples are only derived for the clustering analysis (\ie, the ``LSS analysis'' in SRD). We assume that the lens samples will have tighter priors than the source samples because they are limited to lower redshifts ($z_{\rm lens}<1.2$ as in SRD). Therefore, we take the same priors for lens samples as those for the source samples to be conservative.}. For the alternative sample choice studied in Section~\ref{sec:1sample}, where we assume the lens sample to be the same as the source sample (``lens=source''), the photo-z parameters reduce to only 11.

\paragraph*{Linear galaxy bias}
We assume one linear bias parameter $b_{\rm g}^i$ per lens redshift bin. Similar to the DESC-SRD, the fiducial values follow the simple relation: $b_{\rm g}^i=0.95/G(\langle z^i\rangle)$, where $G(z)$ is the growth function. The total of 10 linear bias parameters are independently marginalised over with a conservative flat prior $[0.4,3.0]$. For the ``lens=source'' choice discussed in Section~\ref{sec:1sample}, we use an even more conservative flat prior $[0.4,5.0]$ since the galaxies have higher redshifts, hence higher $b_{\rm g}^i$. We note that this model may be oversimplified for going into $k_{\rm max}=$0.3$h/$Mpc, especially for the ``lens=source'' case. More complex models may lead to less constraining power of the galaxy survey and enhance the importance of the CMB lensing information (see Section~\ref{sec:conclu} for discussion).

\paragraph*{Multiplicative shear calibration}
We assume one parameter $m^i$ per source redshift bin, which affects $\kappa_{\rm g}$-$\kappa_{\rm g}$, $\delta_{\rm g}$-$\kappa_{\rm g}$, and $\kappa_{\rm g}$-$\kappa_{\rm CMB}$ via
\begin{equation}
    q_{\kappa_{\rm g}}^i\rightarrow (1+m^i)q_{\kappa_{\rm g}}^i
\end{equation}
The total of 10 $m^i$ parameters are independently marginalised over with Gaussian priors. The priors for Y1 are chosen to be consistent with the Y1 requirements in DESC-SRD, while we assume the SRD Y10 requirements for our Y6 analysis.

\paragraph*{Intrinsic alignment (IA)}
Galaxy shapes and orientations are correlated with the underlying density and tidal fields, introducing possible correlations between measured ellipticities. We adopt the ``nonlinear linear alignment'' model \citep{2004PhRvD..70f3526H,2007NJPh....9..444B}, which considers only the ``linear'' response of the elliptical (red) galaxies' shapes to the tidal field sourced by the underlying ``nonlinear'' matter density field (instead of the linear density field, as in \citealp{2001MNRAS.320L...7C}). This model captures most of the IA signal and we do not consider higher-order tidal alignment, tidal torquing models \citep[see \eg,][]{2015JCAP...08..015B,2019PhRvD.100j3506B}, or more complicated IA modelling as a function of galaxy color \citep{2019MNRAS.489.5453S}.

Our implementation follows \cite{2016MNRAS.456..207K} for cosmic shear and \cite{2017MNRAS.470.2100K} for galaxy-galaxy lensing, and we further extend it to $\kappa_{\rm g}$-$\kappa_{\rm CMB}$ cross power spectra. Using the notation in Section~\ref{sssec:probe-modeling}, we can encapsulate the effect as
\begin{equation}
    q_{\kappa_{\rm g}}^i\rightarrow q_{\kappa_{\rm g}}^i + q_I^i~,
\end{equation}
where
\begin{equation}
    q_I^i = -A(m_{\rm lim},z)f_{\rm red}(m_{\rm lim},z)\frac{n_{\rm source}^i(z(\chi))}{\bar{n}_{\rm source}^i}\frac{dz}{d\chi}~.
\end{equation}
$f_{\rm red}(m_{\rm lim},z)$ is the fraction of red galaxies at redshift $z$ evaluated from the GAMA luminosity function \citep{2012MNRAS.420.1239L} assuming a limiting magnitude $m_{\rm lim}=25.3$. $A(m_{\rm lim},z)$ is the IA amplitude at a given limiting magnitude $m_{\rm lim}$ and redshift $z$, computed using the GAMA luminosity function
\begin{equation}
    A(m_{\rm lim},z) = \langle A(L,z)\rangle_{\phi_{\rm red}}\times \left[\Theta(z_1-z)+\Theta(z-z_1)\left(\frac{1+z}{1+z_1}\right)^{\eta_{\rm IA}^{\rm high-z}}\right]~,
\end{equation}
where $\Theta$ is the step function and $\langle\cdots\rangle_{\phi_{\rm red}}$ denotes the average weighted by the luminosity function of red galaxies (see Eq.~24 of \citealp{2016MNRAS.456..207K}). At $z\leq z_1=0.7$, the fiducial redshift scaling is based on the MegaZ-LRG + SDSS LRG sample, \ie,
\begin{equation}
    A(L,z)=\frac{C_1\rho_{\rm cr}}{G(z)}A_{\rm IA}\left(\frac{L}{L_0}\right)^{\beta_{\rm IA}} \left(\frac{1+z}{1+z_0}\right)^{\eta_{\rm IA}}~,
\end{equation}
where $C_1\rho_{\rm cr}=0.0134$ is derived from SuperCOSMOS observations \citep{2004PhRvD..70f3526H,2007NJPh....9..444B}. We adopt the constraints from \cite{2011A&A...527A..26J} with fixed pivot redshift $z_0=0.3$ and pivot luminosity $L_0$ corresponding to an absolute $r$-band magnitude of $-22$. The fiducial values and priors for the nuisance parameters $A_{\rm IA},\beta_{\rm IA},\eta_{\rm IA}$ are given in Table \ref{tab:params}. Since our galaxy sample extends to much higher redshifts, for $z>z_1=0.7$, we extrapolate the fiducial form, but introduce an additional power law scaling with the additional uncertainty parameterised as $\eta_{\rm IA}^{\rm high-z}$. We neglect additional uncertainties in the luminosity function, which can be significant and are discussed in \cite{2016MNRAS.456..207K}.

Together with the multiplicative shear calibration and the MG, $q_{\kappa_{\rm g}}^i$ is altered as
\begin{equation}
    q_{\kappa_{\rm g}}^i(\chi)\rightarrow (1+m^i)[(1+\Sigma(\chi))q_{\kappa_{\rm g}}^i(\chi)+q_I^i(\chi)]~.
\end{equation}

\paragraph*{Baryonic physics}
Baryonic physics impacts the modelling of all the probes via the matter power spectrum on small scales. We mitigate the impact by applying conservative $\ell$ cuts (derived from $k_{\rm max}=0.3h/$Mpc as in Section~\ref{sssec:probe-modeling}) for each $\delta_{\rm g}$-$\delta_{\rm g}$, $\delta_{\rm g}$-$\kappa_{\rm g}$, $\delta_{\rm g}$-$\kappa_{\rm CMB}$ spectra.

We adopt the principal component (PC) decomposition approach \citep{2015MNRAS.454.2451E,2019MNRAS.488.1652H} to account for baryonic uncertainties in the cosmic shear observable. Recently, this method has been applied in the DES Y1 3\x2pt re-analysis to extend the cosmic shear observables down to 2.5 arcmin \citep{2021MNRAS.502.6010H} improving constraints on $S_8$ by $\sim$20\%. We adopt this mitigation strategy for our 6\x2pt analysis as summarised below.

We first compute the difference of dark-matter-only (DMO) to baryonic 6\x2pt data vectors (assuming the same fiducial cosmological and nuisance parameters) for 6 different baryonic scenarios extracted from hydro-simulations: MassiveBlack-II \citep{2015MNRAS.450.1349K,2015MNRAS.453..469T}, Illustris \citep{2014MNRAS.444.1518V,2014MNRAS.445..175G}, Horizon-AGN \citep{2014MNRAS.444.1453D}, Eagle \citep{2015MNRAS.446..521S}, the OWLS-AGN \citep{2010MNRAS.402.1536S,2011MNRAS.415.3649V}, and IllustrisTNG \citep{2018MNRAS.475..676S,2018MNRAS.475..648P,2018MNRAS.475..624N,2018MNRAS.480.5113M,2018MNRAS.477.1206N} (see \citet{2019MNRAS.488.1652H} for a summary of the simulations and motivation of this choice). 

We only contaminate the $\kappa_{\rm g}$-$\kappa_{\rm g}$, $\kappa_{\rm g}$-$\kappa_{\rm CMB}$, and $\kappa_{\rm CMB}$-$\kappa_{\rm CMB}$ components of the data vectors with the baryonic scenarios. These ``difference vectors'' are combined into a ``difference matrix'' in the form of
\begin{equation}
    \bm{\Delta}=\left[\bv{B}_1-\bv{M}\quad \cdots\quad \bv{B}_6-\bv{M}\right]_{N_{\rm D}\times 6}~,
\end{equation}
where $\bv{B}_n$ is the partially contaminated 6\x2pt data vector in the $n$-th baryonic scenario, $\bv{M}$ is the DMO data vector, and $N_{\rm D}$ is the dimension of the data vector. Note that we have excluded the data points beyond the $\ell$ cuts from the data vector.

We weigh this difference matrix with respect to the statistical uncertainties given by the covariance matrix $\bmat{C}$. Writing $\bmat{C}$ in the form of a Cholesky decomposition $\bmat{C}=\bmat{L}\bdot\bmat{L}^\btop$, we can reweigh the difference vectors and decorrelate their elements in the data vector space by applying $\bmat{L}^{-1}$ on each data vector. The difference matrix becomes
\begin{equation}
    \bm{\Delta}_{\rm ch} = \bmat{L}^{-1}\bdot\bm{\Delta}=\bmat{U}_{\rm ch}\bdot\bm{\Sigma}_{\rm ch}\bdot\bmat{V}_{\rm ch}^{\btop}~,
\end{equation}
where we perform a singular value decomposition in the last step. $\bmat{U}_{\rm ch}$ and $\bmat{V}_{\rm ch}$ are square unitary matrices with dimensions $N_{\rm D}\times N_{\rm D}$ and $6\times 6$, respectively. $\bm{\Sigma}_{\rm ch}$ is an $N_{\rm D}\times 6$ rectangular diagonal matrix with the singular values populating the diagonal in descending order. The first 6 columns of the $\bmat{U}_{\rm ch}$ matrix form a set of PC bases, $\bv{v}_{{\rm PC},i}$. For a baryonic scenario $n$, we have
\begin{equation}
    \bmat{L}^{-1}\bdot(\bv{B}_n-\bv{M})=\sum_{i=1}^{6}Q_i\,\bv{v}_{{\rm PC},i}~,
\end{equation}
where $Q_i$ are the PC amplitudes. Using $N_{\rm PC}$ PC modes ($N_{\rm PC}\leq 6$), we can simulate possible baryonic behaviours as
\begin{equation}
    \bv{M}_{\rm bary}(\bv{p},\bv{Q})= \bv{M}_{\rm bary}(\bv{p})+\sum_{i=1}^{N_{\rm PC}}Q_i\,\bmat{L}\bdot\bv{v}_{{\rm PC},i}~,
\end{equation}
where $\bv{Q}=(Q_1,\cdots,Q_{N_{\rm PC}})$.

In our analysis, we take the first $N_{\rm PC}=3$ PCs with amplitudes $Q_{1\sim 3}$, whose fiducial values are set to be zero. $Q_i$'s are marginalised over using conservative Gaussian priors, such that the 1$\sigma$ region of the prior corresponds to half the amplitude of $Q$'s needed to capture the Illustris (not TNG) simulation, as it has a very strong feedback scenario which is highly unlikely given present observations \citep{2016MNRAS.457.3024H}. 

Note that the PCs mostly account for baryonic effects in small-scale data points of $\kappa_{\rm g}$-$\kappa_{\rm g}$. When fixing $\kappa_{\rm g}$-$\kappa_{\rm CMB}$ and $\kappa_{\rm CMB}$-$\kappa_{\rm CMB}$ to be DMO, the $Q_i$ values for all 6 scenarios only change by less than 1\%. Thus, we use the same $Q_i$ priors for both the 6\x2pt and 3\x2pt analyses of a given survey setup (Y1 or Y6).

An important future avenue to explore is the idea that baryonic physics can also be constrained by CMB experiments with the Sunyaev-Zel'dovich (SZ) effect. \cite{Amodeo21} have first demonstrated the direct calibration of baryonic effects in the lensing of BOSS CMASS galaxies using kinetic SZ (kSZ) from ACT measured in \cite{Schaan21}.
The power of SZ measurements is expected to increase rapidly: \cite{Battaglia17} show that the signal-to-noise ratio (SNR) in kSZ with SO (and DESI) will be $\sim100$, making the SZ probe a very promising addition to multi-probe joint dataset analyses in the future.

\subsection{Covariances}\label{ssec:cov}
The Fourier 6\x2pt covariance matrix includes the Gaussian part $\cov^{\rm G}(C(\ell_1),C(\ell_2))$ \citep{2004PhRvD..70d3009H}, the non-Gaussian part from connected 4-point functions in the absence of survey window effect $\cov^{\rm NG,0}(C(\ell_1),C(\ell_2))$ \citep[\eg,][]{2002PhR...372....1C,2009MNRAS.395.2065T}, and the super-sample covariance $\cov^{\rm SSC}(C(\ell_1),C(\ell_2))$ \citep{2013PhRvD..87l3504T}, \ie,
\begin{align}
    \cov(C(\ell_1),C(\ell_2))= &\cov^{\rm G}(C(\ell_1),C(\ell_2))+ \cov^{\rm NG,0}(C(\ell_1),C(\ell_2))\nonumber\\
    &+ \cov^{\rm SSC}(C(\ell_1),C(\ell_2))~.
\end{align}
The modelling and implementation of its 3\x2pt submatrix are described in appendix A of \cite{2017MNRAS.470.2100K}, and the formalism is easily extended to include $\kappa_{\rm CMB}$. The Gaussian covariance includes the probe-specific shot noise terms, \ie,
\begin{align}
    &\cov^{\rm G}(C_{\rm AB}^{ij}(\ell_1),C_{\rm CD}^{kl}(\ell_2))\nonumber\\
    &=\frac{\delta_{\ell_1\ell_2}}{f_{\rm sky}(2\ell_1+1)\Delta\ell_1}\left[\hat{C}_{\rm AC}^{ik}(\ell_1)\hat{C}_{\rm BD}^{jl}(\ell_2)+\hat{C}_{\rm AD}^{il}(\ell_1)\hat{C}_{\rm BC}^{jk}(\ell_2)\right]~,
\end{align}
where $\delta_{ij}$ is the Kronecker delta function, $f_{\rm sky}=\Omega_{\rm s}/(4\pi)$ is the survey's sky coverage fraction, $\Delta\ell$ is the $\ell$ bin width, $\hat{C}_{\rm AC}^{ik}(\ell_1)=C_{\rm AC}^{ik}(\ell_1)+\delta_{ik}\delta_{\rm AC}N_A^i(\ell_1)$. The noise term $N_A^i(\ell)$ are given by
\begin{equation}
    N_{\kappa_{\rm g}}^i(\ell)=\sigma_{\epsilon}^2/\bar{n}_{\rm source}^i~,~~N_{\delta_{\rm g}}^i(\ell)=1/\bar{n}_{\rm lens}^i~,
\end{equation}
and $N_{\kappa_{\rm CMB}}(\ell)$ is the CMB lensing reconstruction noise. $\sigma_\epsilon=0.26$ is the shape noise per component given by the DESC-SRD.
\begin{figure*}
    \centering
    \includegraphics[width=0.45\textwidth]{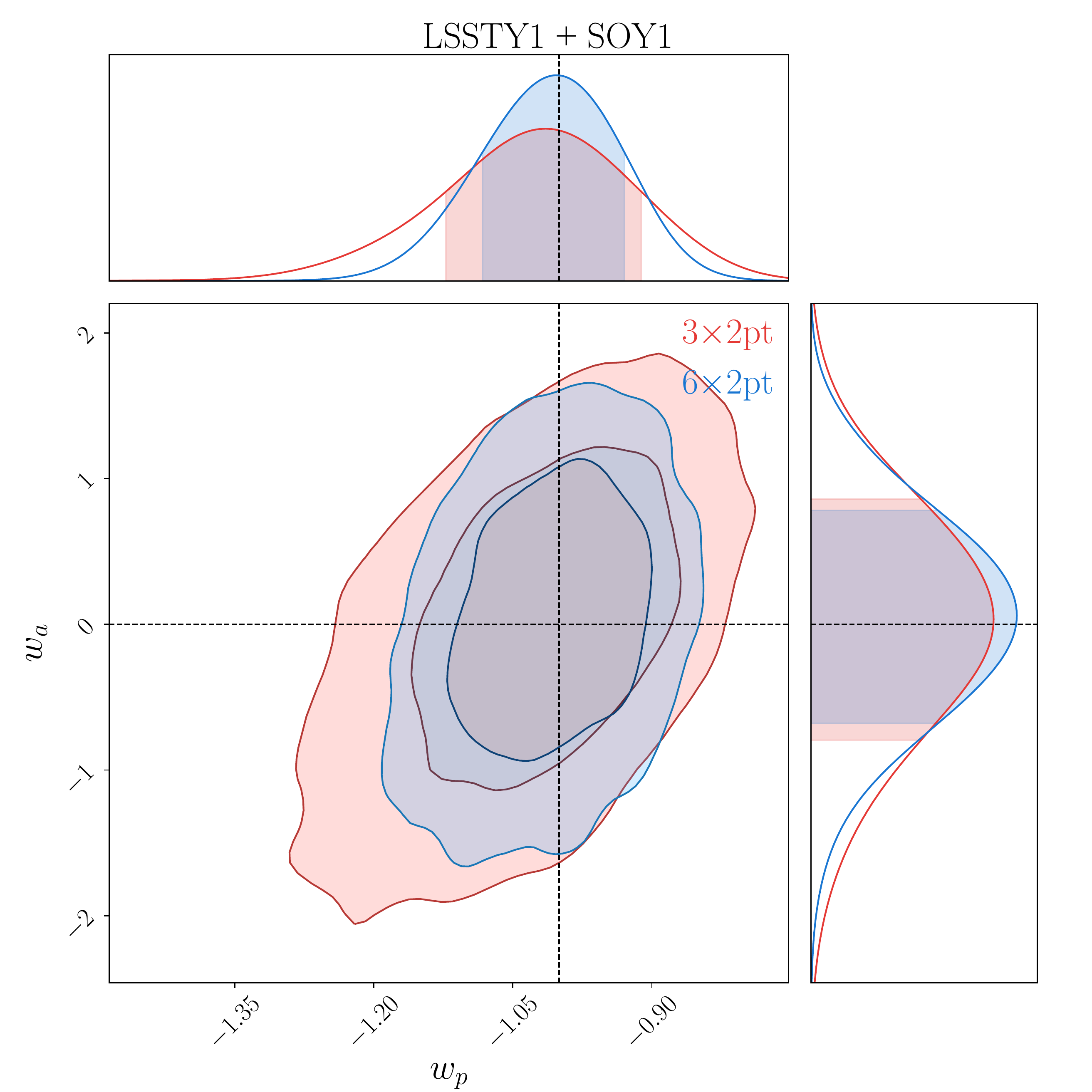}
    \includegraphics[width=0.45\textwidth]{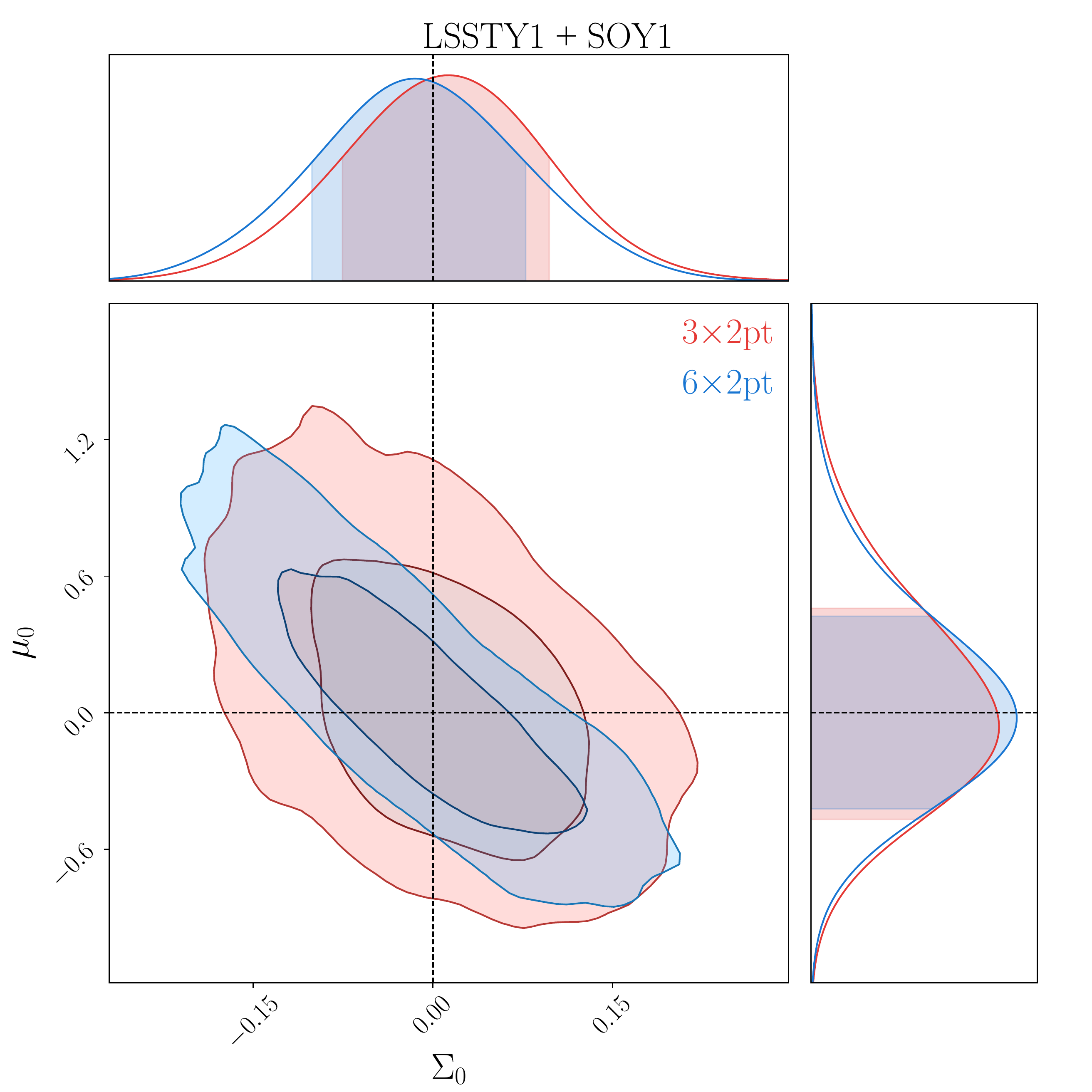}
    \includegraphics[width=0.45\textwidth]{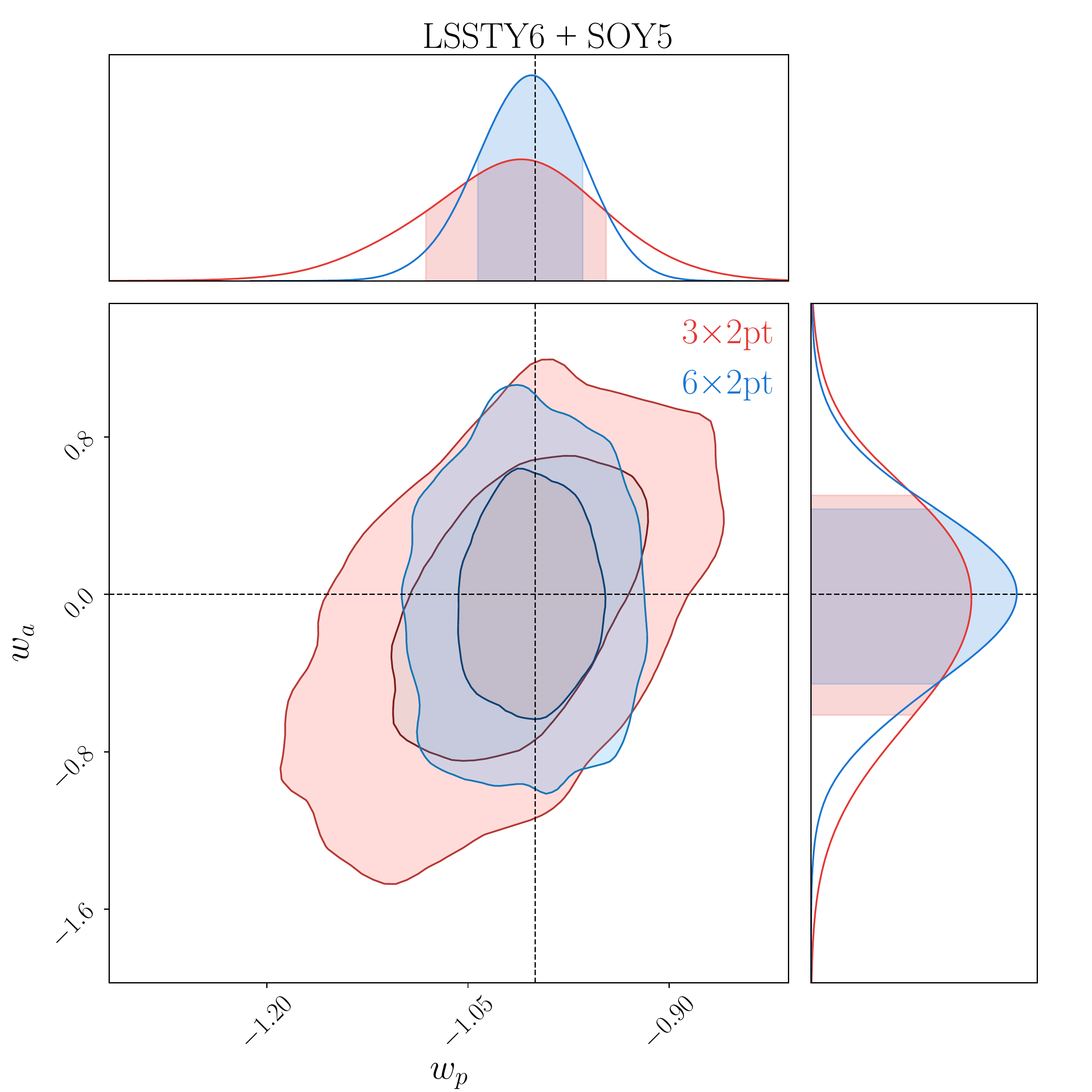}
    \includegraphics[width=0.45\textwidth]{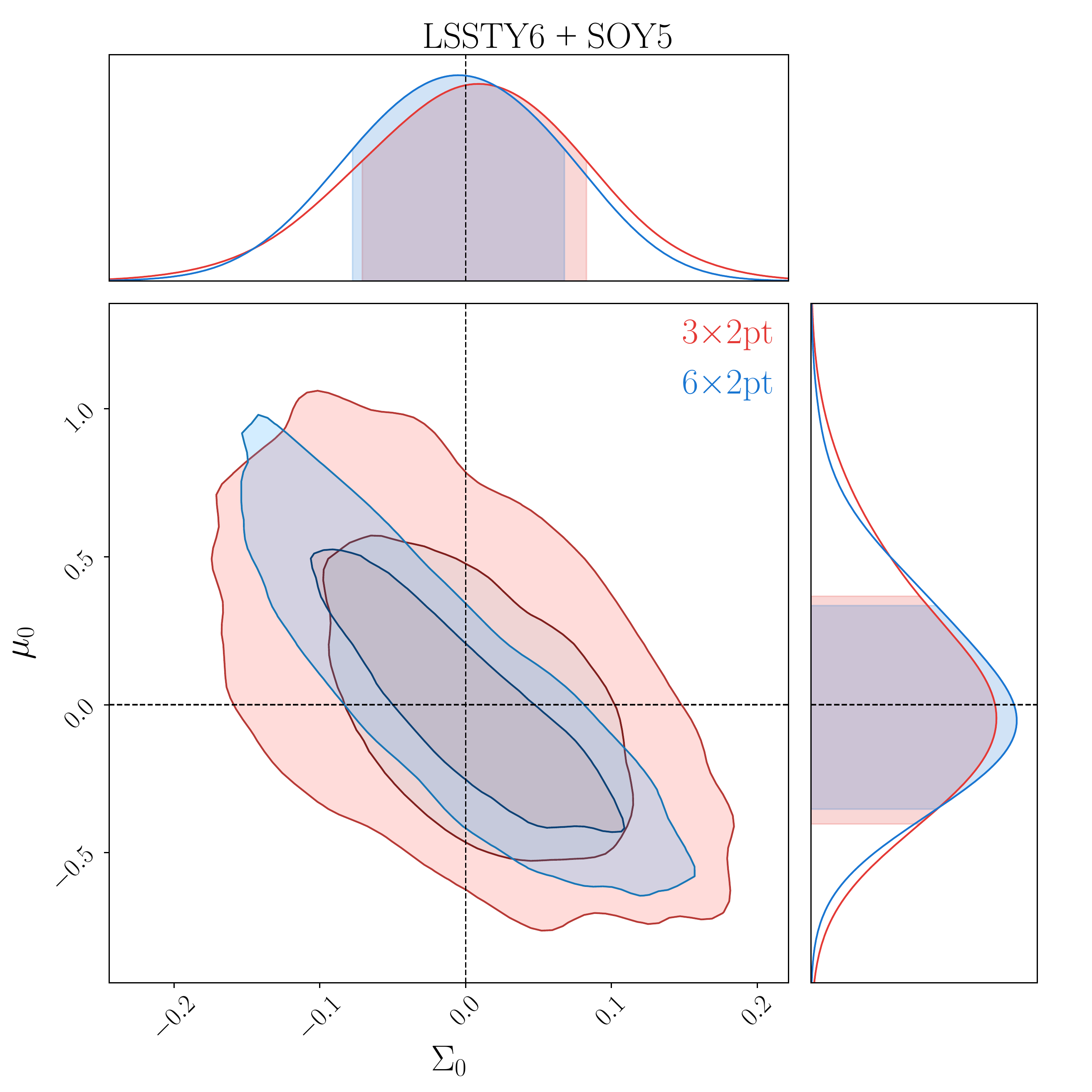}
    \caption{$(w_p,w_a)$ constraints with $w_0-w_a$CDM and $(\mu_0,\Sigma_0)$ constraints with the MG model, from LSST-only 3\x2pt to LSST+SO 6\x2pt probes, for both Y1 (upper panels) and Y6 (lower panels). The improvements in constraining power are quantitatively described in Section~\ref{ssec:3-6x2pt}.}
    \label{fig:3-6x2pt}
\end{figure*}

We estimate the reconstruction noise for CMB lensing from SO Y1 and Y5 using the public SO v3.1 noise calculator\footnote{\url{https://github.com/simonsobs/so_noise_models}} \citep{Ade19}.

For SO Y5, we simply use the file \path{nlkk_v3_1_0deproj0_SENS1_fsky0p4_qe_lT30-3000_lP30-5000.dat}.
This corresponds to the ``baseline'' SO configuration,
where the lensing reconstruction is performed using the minimum variance quadratic estimator (QE) for CMB lensing \citep{2002ApJ...574..566H}.
This lensing estimator relies on the internal linear combination (ILC) CMB maps in temperature and polarisation, with $\ell_\text{min, T} = \ell_\text{min, P} = 30$, $\ell_\text{max, T}=3000$ and $\ell_\text{max, P}=5000$.

Since no official SO Y1 lensing noise curve exists, we estimate it in two approximate ways, and check that these two methods agree within $\sim$10\%.
We assume that the SO Y1 and Y5 footprints are identical and that most of the lensing information in SO Y1 comes from temperature, rather than polarisation. 

In the first approximate method, we use the SO noise calculator in the ``baseline'' configuration with a sky coverage $f_\text{sky}=0.4$, a typical elevation of 50~deg, and one year of observations.
We then combine the 93 and 145~GHz temperature maps with inverse-variance detector noise weighting.
We then add the fiducial level of foreground power at 145~GHz from \cite{Dunkley13}.
This provides an approximation to the Y1 ILC noise and foreground levels.
We then run the code \texttt{LensQuest}\footnote{\url{https://github.com/EmmanuelSchaan/LensQuEst}} \citep{Schaan19} to obtain the lensing noise from this approximate ILC temperature map.

In the second approximate method, we start with the Y5 TT-only CMB lensing noise curve, which uses the exact Y5 ILC temperature map.
We then scale it up by a factor 1.35 to simulate SO Y1.
Indeed, the low lensing-$L$ plateau of the lensing noise is inversely proportional to the number of signal-dominated Fourier modes in the CMB temperature map.
The scaling factor in the lensing noise from Y5 to Y1 is thus simply the ratio of the numbers of signal-dominated modes for Y5 and Y1.
To compute it, we use again the noise calculator, in the ``baseline'' configuration, with a sky coverage $f_\text{sky}=0.4$ and a typical elevation of 50~deg.
As above, we simply combine the 93 and 145~GHz temperature maps with inverse-detector-noise-variance weighting.
This yields the approximate ILC noise levels for 1 and 5 years.
We find that Y5 temperature has 1.35 times more signal-dominated Fourier modes than Y1 temperature.

We verify that the two approximations to the Y1 lensing noise agree to within $\sim10\%$ on all scales.
We also verify that the \texttt{LensQuest} code reproduces the public Y5 lensing noise curves.

In what follows, we assume the CMB lensing to be free of any systematic effect.
In practice, extragalactic foregrounds in temperature can be a limiting factor for SO \citep{vanEngelen14, Ferraro18, Schaan19}.
However, modified estimators, such as the shear-only QE \citep{Schaan19}, the point-source-hardened \citep{Namikawa13, Osborne14, Sailer20} or the profile-hardened QE \citep{Sailer20} or the gradient-cleaned estimators \citep{Madhavacheril18, Darwish21}, have been shown to effectively mitigate these foregrounds in simulation, at virtually no cost in signal to noise.
Furthermore, these foregrounds are likely only significant in temperature, rather than polarisation.
They are thus less concerning for SO Y5 than Y1.

\section{Simulated Likelihood Analysis Results}\label{sec:forecast}
We use \texttt{emcee} \citep{2013PASP..125..306F} to sample the parameter space. For each analysis chain, we use 1128 random walkers running in parallel, 8000 steps each, hence 9.024M steps per chain. We remove 70\% of the steps as burn-in and check the convergence for each chain.

\subsection{From 3\x2pt to 6\x2pt: Cosmology and systematics constraints}\label{ssec:3-6x2pt}
The addition of CMB lensing from SO allows us to access cosmological information at higher redshifts compared to using data from LSST only, and it improves constraints on systematic effects through self-calibration.

In Figure \ref{fig:3-6x2pt}, we show the results from our 3\x2pt and 6\x2pt simulated analyses on the dark energy equation-of-state parameters $(w_p, w_a)$, where $w_p$ is $w(a)$ evaluated at the pivot redshift $z_p=0.5$, and the MG parameters $(\mu_0,\Sigma_0)$.

We quantify the improvement in the dark energy parameters through the commonly used ``Figure-of-Merit'' (FoM) in the $(w_0, w_a)$ plane, ${\rm FoM}=[\det\,\cov(w_0,w_a)]^{-1/2}$ \citep{2001PhRvD..64l3527H,2006astro.ph..9591A},\footnote{The FoM here is larger by a factor of $\pi$ than that in \cite{2006astro.ph..9591A} where the FoM is defined as the inverse area of the posterior 1$\sigma$ ellipse.} where the parameter covariance is estimated from the posteriors of $w_0$ and $w_a$. 

We find that in Y1, combining LSST and SO probes improves FoM by 53\% from 15 (LSST-only 3\x2pt) to 23 (6\x2pt), and in LSST Y6 + SO Y5, FoM is improved by 92\% from 36 (LSST-only 3\x2pt) to 69 (6\x2pt).

The FoM definition can be applied to any parameter plane of interest $(p_1,p_2)$ as ${\rm FoM}_{p_1,p_2}=[\det\,\cov(p_1,p_2)]^{-1/2}$. In addition to dark energy, we are interested in the constraining power on modified gravity and find that going from 3\x2pt to 6\x2pt, in the $(\mu_0,\Sigma_0)$ plane, the ${\rm FoM}_{\mu_0,\Sigma_0}$ increases by 72\% in Y1 and 106\% in Y6.

\begin{figure}
    \centering
    \includegraphics[width=\columnwidth]{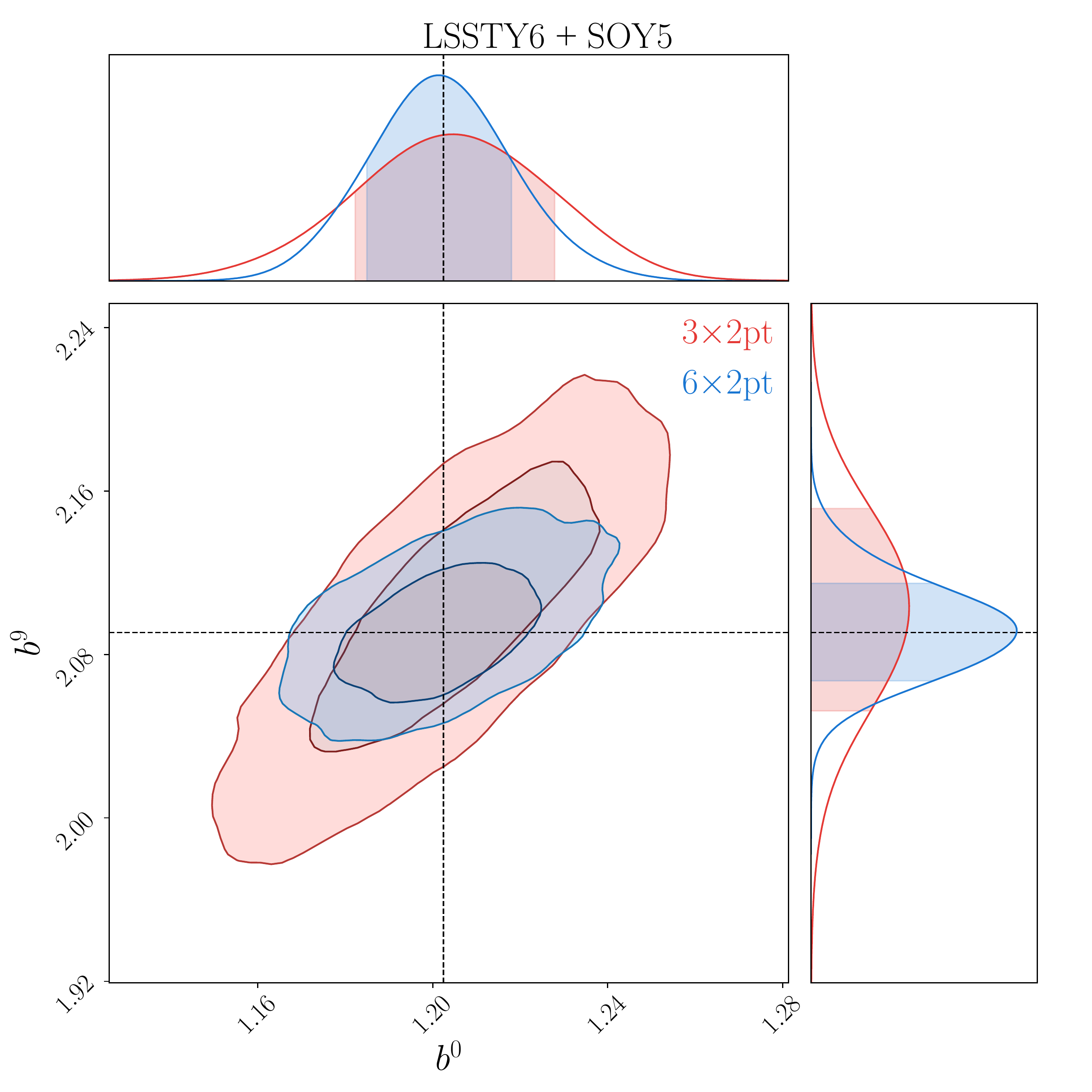}
    \caption{Linear galaxy bias parameter ($b^0,b^9$) constraints from LSST-only 3\x2pt to LSST+SO 6\x2pt probes for Y6, assuming $w_0-w_a$CDM model. The improvements in constraining power of bias parameters in other tomographic bins are similar and are quantitatively described in Section~\ref{ssec:3-6x2pt}.}
    \label{fig:3-6x2pt_gbias}
\end{figure}

We further explore the gain in information on important nuisance parameters. Figure \ref{fig:3-6x2pt_gbias} shows the 3\x2pt and 6\x2pt constraints on the galaxy bias parameters $(b^0,b^9)$. Other parameters show similar improvements as presented in Table~\ref{tab:param-1d}. For Y1, the 1$\sigma$ constraints improve from 25\% (bin 0) to $\sim 48\%$ (bin 8 and 9). For Y6, the improvement increases from 29\% to 51\%.

\subsection{From Y1 to Y6: Increasing depth and survey area}\label{ssec:y1-6}
From Y1 to Y6, LSST is expected to increase its depth as well as its survey area. The former leads to an increase in the number density of galaxies, hence a decrease in the shot/shape noise, while the latter leads to an overall reduction of cosmic variance.

\begin{figure*}
    \centering
    \includegraphics[width=0.45\textwidth]{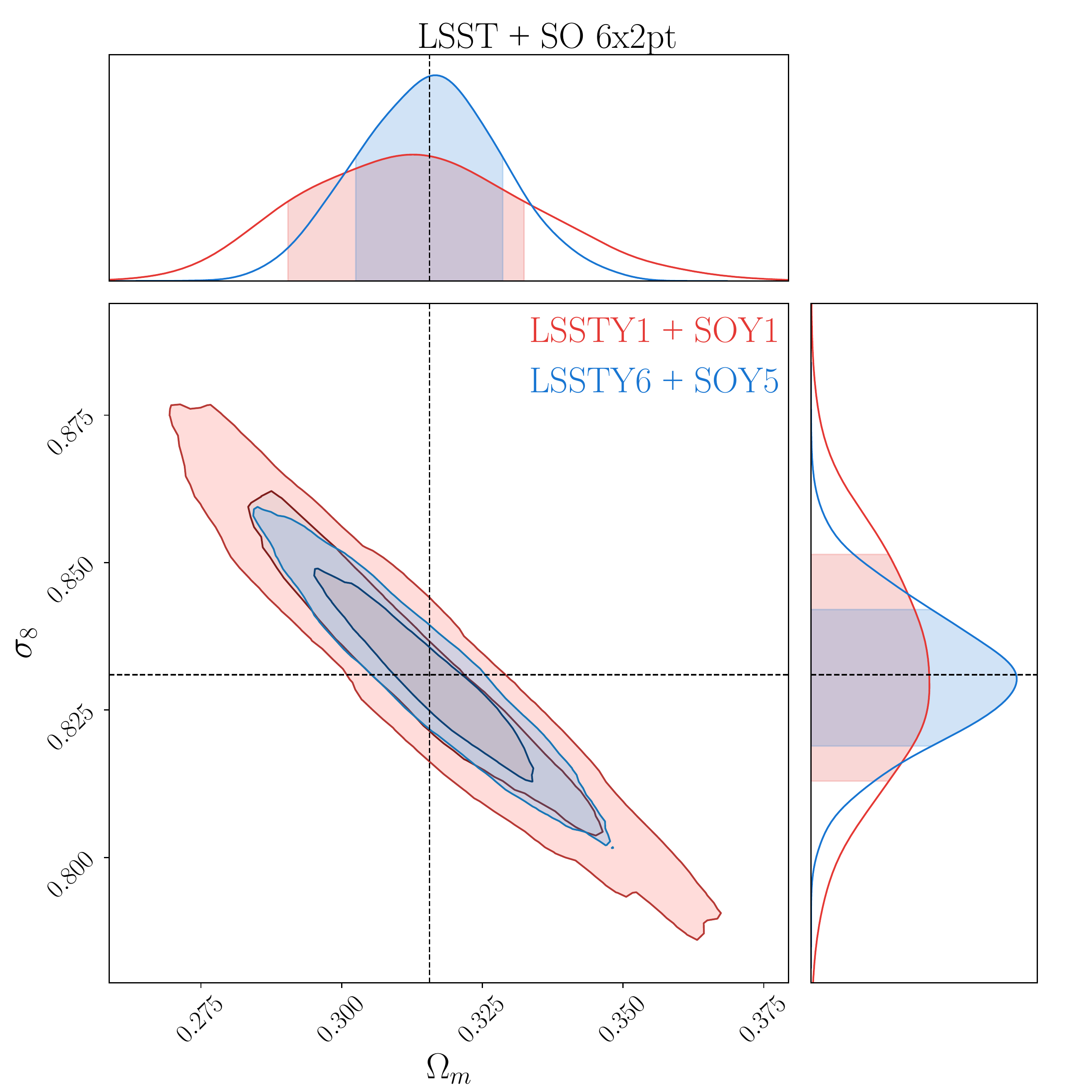}
    \includegraphics[width=0.45\textwidth]{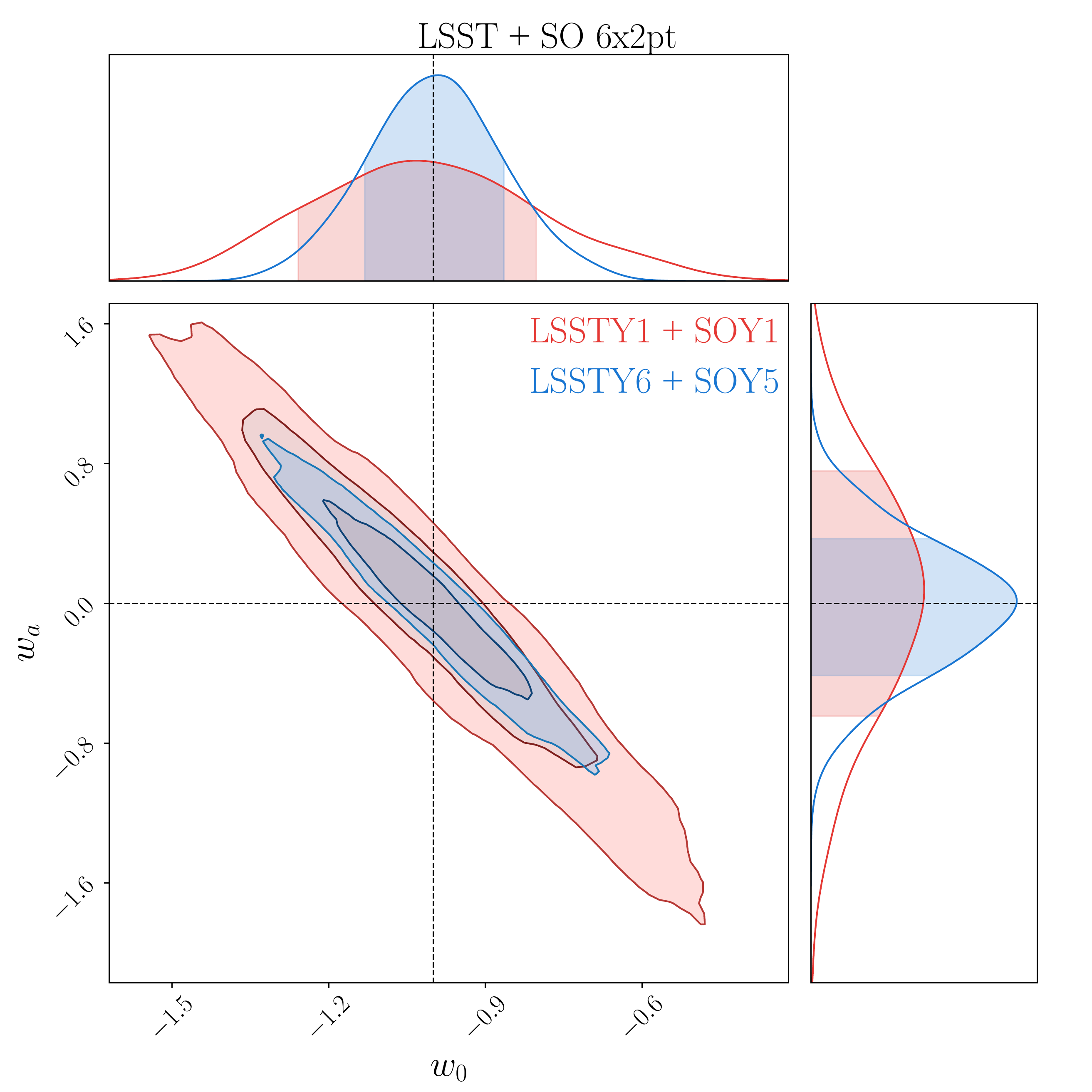}
    \caption{LSST+SO 6\x2pt $(\Omega_m,\sigma_8)$ and $(w_p,w_a)$ constraints assuming $w_0-w_a$CDM model for Y1 (left panel) and Y6 (right panel). The improvements in constraining power are quantitatively described in Section~\ref{ssec:y1-6}.}
    \label{fig:y1-6}
\end{figure*}

In Figure \ref{fig:y1-6}, we compare the $\Omega_m-\sigma_8$ and $w_0-w_a$ constraints from 6\x2pt between Year 1 and Year 6. The FoM in the $w_0-w_a$ plane triples from 23 to 69.

\begin{table}
    \centering
    \begin{tabular}{|c|c|c|c|c|}
    \hline
         & Y1 3\x2pt & Y1 6\x2pt & Y6 3\x2pt & Y6 6\x2pt \\
         \hline
     SRD & 311 & 319 & 429 & 444 \\
     ``lens=source'' & 406 & 412 & 608 & 616\\
     \hline
     increased by & 30.5\% & 29.2\% & 41.7\% & 38.7\%\\
         \hline
    \end{tabular}
    \caption{SNRs (after scale cuts) of the data vectors at fiducial cosmological and systematic parameter values. For each case, the SNR is increased by $\sim$30-40\% comparing to the SRD sample case when the ``lens=source'' choice is adopted.}
    \label{tab:snr}
\end{table}

\begin{figure*}
    \centering
    \includegraphics[width=0.7\textwidth]{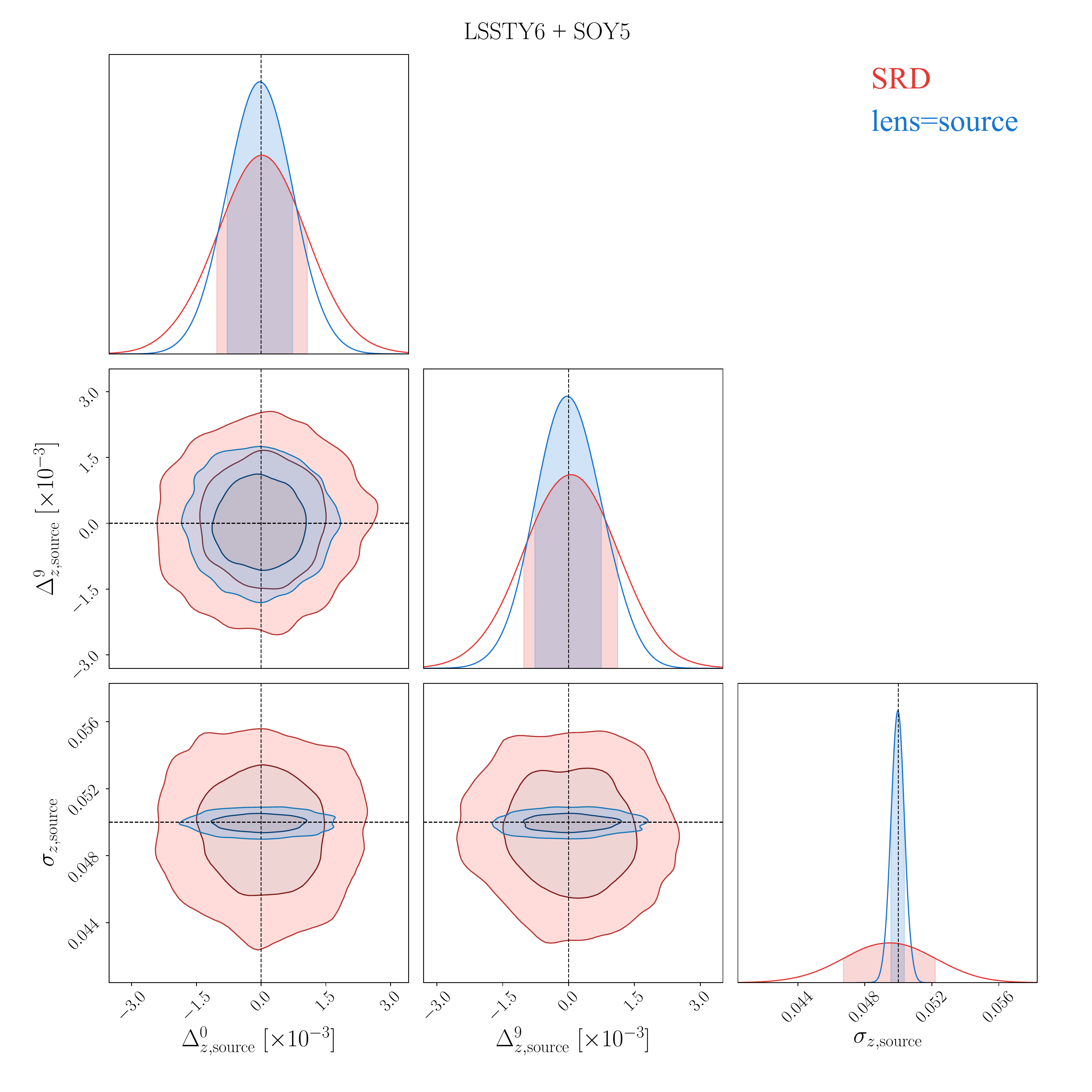}
    \caption{Source sample photo-z shift parameter $(\Delta_{z,\rm source}^0,\Delta_{z,\rm source}^9,\sigma_{z,\rm source})$ constraint with $w_0-w_a$CDM model from LSST Y6 + SO Y5 6\x2pt, adopting the SRD lens sample choice versus the ``lens=source'' choice. The improvements in other shift parameters' calibration are similar and are quantitatively described in Section~\ref{sec:1sample}. The $\sigma_{z,\rm source}$ constraint is significantly improved because it is a shared parameter among all 10 tomographic bins.}
    \label{fig:y6-6x2-1sample-zs}
\end{figure*}

\begin{table*}
    \centering
    Posterior 1D $\sigma$ of cosmological parameters\\
    \centering
    \begin{tabular}{|c|c|c|c|c|c|c|c|c|c||c|c|c|}
    \hline
    \textbf{LSST Y1 + SO Y1} & \multicolumn{9}{ c|| }{$\sigma_{p_i}$ for Fiducial $w_0-w_a$CDM} &\multicolumn{3}{c|}{$\sigma_{p_i}$ for MG}\\
    \hline
    & $\Omega_m$ & $\sigma_8$ & $n_s$ & $w_0$ & $w_a$ & $\Omega_b$ & $h$ & $s_8$ $^*$ & $w_p$ $^*$ & $\mu_0$ & $\Sigma_0$ & $\mu_0+4\Sigma_0$\\
    \hline
    6\x2pt & $0.019$ & $0.018$ & $0.036$ & $0.21$ & $0.65$ & $0.0040$ & $0.054$ & $0.0059$ & \textbf{0.069} & $0.40$ & $0.080$ & \textbf{0.23} \\
    LSST-only 3\x2pt & $0.020$ & $0.020$ & $0.039$ & $0.22$ & $0.76$ & $0.0040$ & $0.059$ & $0.0076$ & \textbf{0.098} & $0.43$ & $0.079$ & \textbf{0.41} \\
    6\x2pt ``lens=source'' & $0.019$ & $0.017$ & $0.035$ & $0.19$ & $0.61$ & $0.0038$ & $0.054$ & $0.0054$ & \textbf{0.056} & - &  - & - \\
    \hline\hline
    \textbf{LSST Y6 + SO Y5} & $\Omega_m$ & $\sigma_8$ & $n_s$ & $w_0$ & $w_a$ & $\Omega_b$ & $h$ & $s_8$ & $w_p$ & $\mu_0$ & $\Sigma_0$ & $\mu_0+4\Sigma_0$\\
    \hline
    6\x2pt & $0.014$ & $0.012$ & $0.029$ & $0.14$ & $0.40$ & $0.0037$ & $0.043$ & $0.0040$ & \textbf{0.036} & $0.32$ & $0.064$ & \textbf{0.16} \\
    LSST-only 3\x2pt & $0.015$ & $0.014$ & $0.033$ & $0.15$ & $0.50$ & $0.0038$ & $0.050$ & $0.0044$ & \textbf{0.062} & $0.35$ & $0.070$ & \textbf{0.31} \\
    6\x2pt ``lens=source'' & $0.013$ & $0.011$ & $0.024$ & $0.12$ & $0.36$ & $0.0037$ & $0.039$ & $0.0036$ & $0.031$ & - &  - & - \\
    \hline
    \end{tabular}\\
    *Define $s_8=\sigma_8(\Omega_m/0.3)^{0.35}$, $w_p=w(1/(1+z_p))$ at $z_p=0.5$.
    \newline
    \newline
    Posterior 1D $\sigma$ of galaxy bias and photo-z parameters\\
    \begin{tabular}{|c|c|c|c|c|c|c|c|c|c|c|c|}
    \hline
    \textbf{LSST Y1 + SO Y1} & \multicolumn{10}{c|}{$10^2\times b^i$, $i=0,1,2,\cdots,9$} & \\
    \hline
    6\x2pt & $2.1$ & $2.1$ & $2.2$ & $2.3$ & $2.2$ & $2.4$ & $2.6$ & $2.6$ & $2.7$ & $3.1$ & \\
    LSST-only 3\x2pt & $2.8$ & $3.1$ & $3.5$ & $3.7$ & $3.8$ & $4.3$ & $4.7$ & $4.8$ & $5.2$ & $5.9$ & \\
    \hline
     & \multicolumn{10}{c|}{$10^4\times \Delta_{z,\rm lens}^i$, $i=0,1,2,\cdots,9$} & $10^4\times\sigma_{z,\rm lens}$ \\
    \hline
    6\x2pt & $19$ & $18$ & $18$ & $18$ & $18$ & $17$ & $19$ & $18$ & $19$ & $19$ & $4.1$ \\
    LSST-only 3\x2pt & $18$ & $18$ & $19$ & $18$ & $17$ & $18$ & $19$ & $18$ & $19$ & $19$ & $4.3$ \\
    \hline
     & \multicolumn{10}{c|}{$10^4\times \Delta_{z,\rm source}^i$, $i=0,1,2,\cdots,9$} & $10^4\times\sigma_{z,\rm source}$ \\
    \hline
    6\x2pt & $19$ & $18$ & $18$ & $18$ & $17$ & $17$ & $18$ & $17$ & $18$ & $19$ & $43$ \\
    LSST-only 3\x2pt & $19$ & $18$ & $18$ & $18$ & $18$ & $17$ & $18$ & $18$ & $18$ & $20$ & $46$ \\
    6\x2pt ``lens=source'' & $14$ & $13$ & $12$ & $13$ & $12$ & $12$ & $12$ & $13$ & $13$ & $14$ & \textbf{6.3} \\
    \hline\hline
    \textbf{LSST Y6 + SO Y5} & \multicolumn{10}{c|}{$10^2\times b^i$, $i=0,1,2,\cdots,9$} & \\
    \hline
    6\x2pt & $1.5$ & $1.5$ & $1.7$ & $1.6$ & $1.7$ & $1.9$ & $1.9$ & $1.8$ & $2.0$ & $2.2$ 
    & \\
    LSST-only 3\x2pt & $2.1$ & $2.3$ & $2.6$ & $2.7$ & $2.9$ & $3.3$ & $3.6$ & $3.5$ & $3.9$ & $4.5$ & \\
    \hline
     & \multicolumn{10}{c|}{$10^4\times \Delta_{z,\rm lens}^i$, $i=0,1,2,\cdots,9$} & $10^4\times\sigma_{z,\rm lens}$ \\
    \hline
    6\x2pt & $9.6$ & $9.5$ & $9.2$ & $9.4$ & $9.5$ & $9.2$ & $9.6$ & $9.5$ & $9.4$ & $9.9$ & $2.7$ \\
    LSST-only 3\x2pt & $9.4$ & $9.6$ & $9.5$ & $9.2$ & $9.4$ & $9.3$ & $9.6$ & $9.2$ & $9.6$ & $9.6$ & $2.7$ \\
    \hline
     & \multicolumn{10}{c|}{$10^4\times \Delta_{z,\rm source}^i$, $i=0,1,2,\cdots,9$} & $10^4\times\sigma_{z,\rm source}$ \\
    \hline
    6\x2pt & $9.6$ & $8.5$ & $8.0$ & $8.0$ & $8.4$ & $8.4$ & $8.2$ & $8.4$ & $8.7$ & $9.7$ & $25$ \\
    LSST-only 3\x2pt & $9.4$ & $8.3$ & $8.2$ & $8.1$ & $8.5$ & $8.5$ & $8.2$ & $8.3$ & $8.9$ & $9.9$ & $25$ \\
    6\x2pt ``lens=source'' & $6.9$ & $6.4$ & $6.3$ & $6.1$ & $6.0$ & $6.0$ & $6.1$ & $6.3$ & $6.4$ & $7.0$ & \textbf{3.7} \\
    \hline
    \end{tabular}
    \newline
    \newline
    Posterior 1D $\sigma$ of shear calibration, IA, and baryon parameters\\
    \begin{tabular}{|c|c|c|c|c|c|c|c|c|c|c|}
    \hline
    \textbf{LSST Y1 + SO Y1} & \multicolumn{10}{c|}{$10^3\times m^i$, $i=0,1,2,\cdots,9$} \\
    \hline
    6\x2pt & $13$ & $11$ & $9.9$ & $8.9$ & $8.0$ & $7.3$ & $6.7$ & $6.3$ & $5.8$ & $5.7$ \\
    LSST-only 3\x2pt & $12$ & $11$ & $9.8$ & $8.7$ & $7.9$ & $7.1$ & $6.8$ & $6.6$ & $6.6$ & $6.7$ \\
    6\x2pt ``lens=source'' & $12$ & $11$ & $9.4$ & $8.2$ & $7.4$ & $6.8$ & $6.2$ & $5.7$ & $5.6$ & $5.3$ \\
    \hline
     & $A_{\rm IA}$ & $\beta_{\rm IA}$ & $\eta_{\rm IA}$ & $\eta^{\rm high-z}_{\rm IA}$ & $Q_1$ & $Q_2$ & $Q_3$ &  &  & \\
    \hline
    6\x2pt & $2.2$ & $0.55$ & $1.6$ & $0.55$ & $6.0$ & $1.0$ & $0.63$ &  &  & \\
    LSST-only 3\x2pt & $2.3$ & $0.58$ & $1.7$ & $0.56$ & $6.2$ & $1.1$ & $0.65$ &  &  & \\
     6\x2pt ``lens=source'' & $2.1$ & $0.53$ & $1.5$ & $0.56$ & $5.9$ & $1.0$ & $0.70$ &  &  & \\
    \hline\hline
    \textbf{LSST Y6 + SO Y5} & \multicolumn{10}{c|}{$10^3\times m^i$, $i=0,1,2,\cdots,9$} \\
    \hline
    6\x2pt & $3.0$ & $2.8$ & $2.7$ & $2.5$ & $2.4$ & $2.3$ & $2.1$ & $1.9$ & $1.9$ & $1.9$ \\
    LSST-only 3\x2pt & $3.1$ & $2.9$ & $2.7$ & $2.6$ & $2.4$ & $2.3$ & $2.1$ & $2.0$ & $1.9$ & $2.0$ \\
    6\x2pt ``lens=source'' & $3.0$ & $2.8$ & $2.7$ & $2.5$ & $2.3$ & $2.1$ & $2.0$ & $1.9$ & $1.8$ & $1.8$ \\
    \hline
     & $A_{\rm IA}$ & $\beta_{\rm IA}$ & $\eta_{\rm IA}$ & $\eta^{\rm high-z}_{\rm IA}$ & $Q_1$ & $Q_2$ & $Q_3$ &  &  & \\
    \hline
    6\x2pt & $2.1$ & $0.41$ & $1.1$ & $0.57$ & $8.2$ & $1.2$ & $1.3$ &  &  & \\
    LSST-only 3\x2pt & $2.0$ & $0.41$ & $1.1$ & $0.56$ & $9.0$ & $1.3$ & $1.3$ &  &  & \\
     6\x2pt ``lens=source'' & $2.0$ & $0.38$ & $1.0$ & $0.55$ & $7.0$ & $1.1$ & $1.2$ &  &  & \\
    \hline
    \end{tabular}
    \caption{1D posterior constraints for LSST-only 3\x2pt and LSST+SO 6\x2pt SRD and ``lens=source'' cases. We do not list the galaxy bias parameter constraints for the ``lens=source'' since these parameters are for a different sample and cannot be compared to the SRD case.}
    \label{tab:param-1d}
\end{table*}

\subsection{One galaxy sample to rule them all}\label{sec:1sample}
Designing an optimal lens sample is one of the most important analysis choices in 3\x2pt analyses. The idea to increase number density to reduce shot noise and to increase redshift range of the sample is frequently countered with considerations about the redshift accuracy. The source sample however is usually well defined as the sweet spot of maximising the number density for which acceptable shapes can be measured.

The idea of using the source sample as the lens sample (``lens=source''), \ie, using the same galaxies for both clustering and lensing, has the obvious advantage that it eliminates half of the photo-z nuisance parameters. Recently, the advantages of choosing ``lens=source'' have been discussed in \cite{2020JCAP...12..001S}, where the authors show that a significant improvement in photo-z uncertainties can be achieved.

We repeat the 6\x2pt likelihood analysis with the ``lens=source'' sample choice, which eliminates 11 photo-z nuisance parameters from the sampled parameter space. 

We find a significant improvement (21\%-33\% for all bins and both years) in the 10 photo-z shift parameters $\Delta^i_{z, \rm source}$ and a dramatic improvement (a factor of $\sim 7$ in the 1D posterior) in the dispersion parameter $\sigma_{z, \rm source}$, as presented in Table~\ref{tab:param-1d} and illustrated in Figure \ref{fig:y6-6x2-1sample-zs}. We also find mild improvement in most of the cosmological parameter constraints, similar to \cite{2020JCAP...12..001S}. 

We compute the SNRs for both the SRD lens sample choice and the ``lens=source'' choice, for both 3\x2pt and 6\x2pt analyses in Y1 and Y6, see Table~\ref{tab:snr}. We find a $\sim$30\% improvement in Y1 and a $\sim$40\% improvement in Y6 when adopting ``lens=source''. 

Although this does not translate into strong improvements on cosmological parameters in the parameter space considered here, we expect that the increase in constraining power will be more relevant for more complex physics models. We note that the gain in photo-z self-calibration ability reflects that this choice can make the analysis more robust against potential degeneracies in the parameter space that might occur when dealing with actual data. In addition, the focus on one joint sample requires less work in the upstream catalogue generation. 

\section{Catastrophic Photo-z Outliers}\label{sec:outlier}
In the past sections, we have assumed that photo-z uncertainties can be well described as Gaussian with uncertainties that can be parameterised through a shift parameter in the mean per tomographic bin and a parameter changing the width of the distribution. 

Catastrophic photo-z outliers however cannot be described with this approach due to the large discrepancy between estimated and true redshifts of the galaxies. These outliers have been shown to lead significant cosmological parameter biases in weak lensing surveys if they are not accounted for \citep[e.g.,][]{2010MNRAS.401.1399B,2020JCAP...12..001S}. We simulate how significant these biases are for LSST in Section~\ref{ssec:impact}, develop a simple but efficient mitigation scheme in Section~\ref{ssec:outlier_model}, and test its performance in Section~\ref{ssec:mitigation_tests}. 

\subsection{Impact Study}\label{ssec:impact}
In order to study the impact of catastrophic outliers on 3\x2pt cosmology constraints, we first need to calculate a redshift distribution that includes realistic catastrophic photo-z outliers $n_{X,\rm out}^i(z)$, propagate this into a contaminated data vector, and finally run a simulated analysis without mitigating the photo-z outlier effects.

For the first step, we employ the LSST galaxy mock catalogues from \cite{2018AJ....155....1G,2020AJ....159..258G}, which are designed specifically to realistically model and improve photo-z estimates for LSST. Since these mocks have only been simulated for LSST Y2, Y5 and Y10, we take the Y5 mock galaxy catalogue and study the impact in the context of our Y6 analysis design.

\begin{figure}
    \centering
    \includegraphics[width=0.5\textwidth]{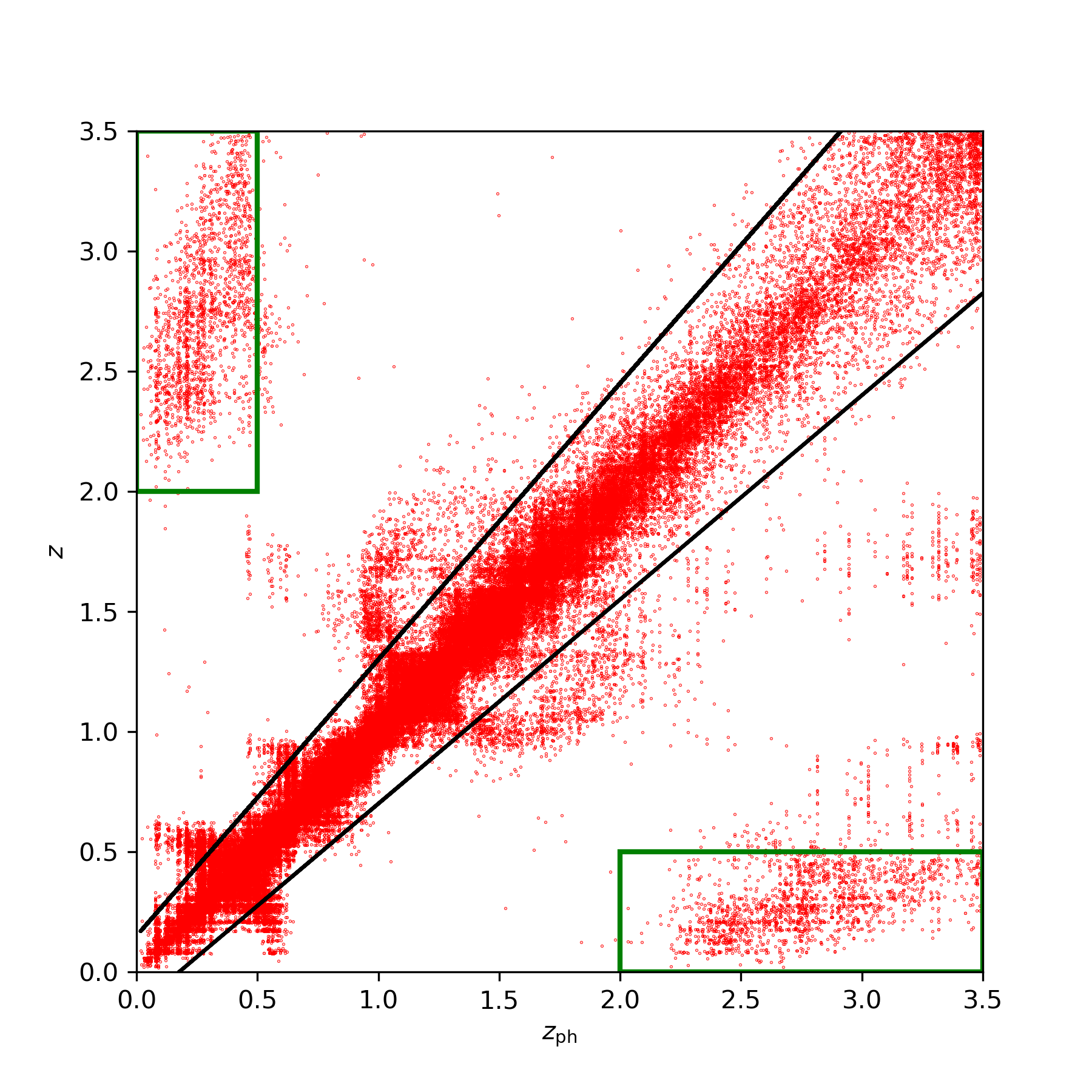}
    \caption{The simulated mock galaxy catalogue for photo-z studies of LSST Year 5 \citep[used in][]{2018AJ....155....1G,2020AJ....159..258G}. The catalogue contains the true redshifts $z$ and the photometric redshifts $z_{\rm ph}$ of 178,269 galaxies. The solid black lines mark $|z-z_{\rm ph}|=3\times 0.05(1+z_{\rm ph})$, within which we assume that photo-z errors are distributed as a Gaussian perfectly described by our fiducial Gaussian photo-z model. Each of the 2 green rectangles encloses a group of catastrophic outliers. We call them ``islands''.}
    \label{fig:z-zph}
\end{figure}

Figure \ref{fig:z-zph} shows the true and photometric redshifts of 178,269 galaxies in the Y5 mock catalogue. Several interesting features can be observed in this figure, which will later guide our development of the mitigation scheme: First, most of galaxies are distributed around the diagonal line $z=z_{\rm ph}$. Second, a significant portion of galaxies fall far off the diagonal. In particular, there are two ``islands'' of galaxies staying at the upper-left and lower-right corners. Third, we note that some galaxies extend from the diagonal and form structures we describe as ``horns''. We assume most of the galaxies around the diagonal with $|z-z_{\rm ph}|<=3\times 0.05(1+z_{\rm ph})$ (\ie, within the two black lines) are well described by the Gaussian photo-z error distribution we assumed in the previous sections. These black lines contain galaxies with photo-z error within 3$\sigma$, assuming $\sigma_z=0.05$. We denote all the other galaxies as outliers. Most of these outliers are due to the Balmer/Lyman break shifting between filters, making certain colours degenerate at certain redshifts. 

We use the outlier galaxies only to calculate the probability matrix of a galaxy having true redshift $z$ given its photometric redshift $z_{\rm ph}$, $p_{\rm out}(z|z_{\rm ph})$, and we call this catalogue the ``full-outlier'' catalogue.

\begin{figure*}
    \centering
    \includegraphics[width=0.45\textwidth]{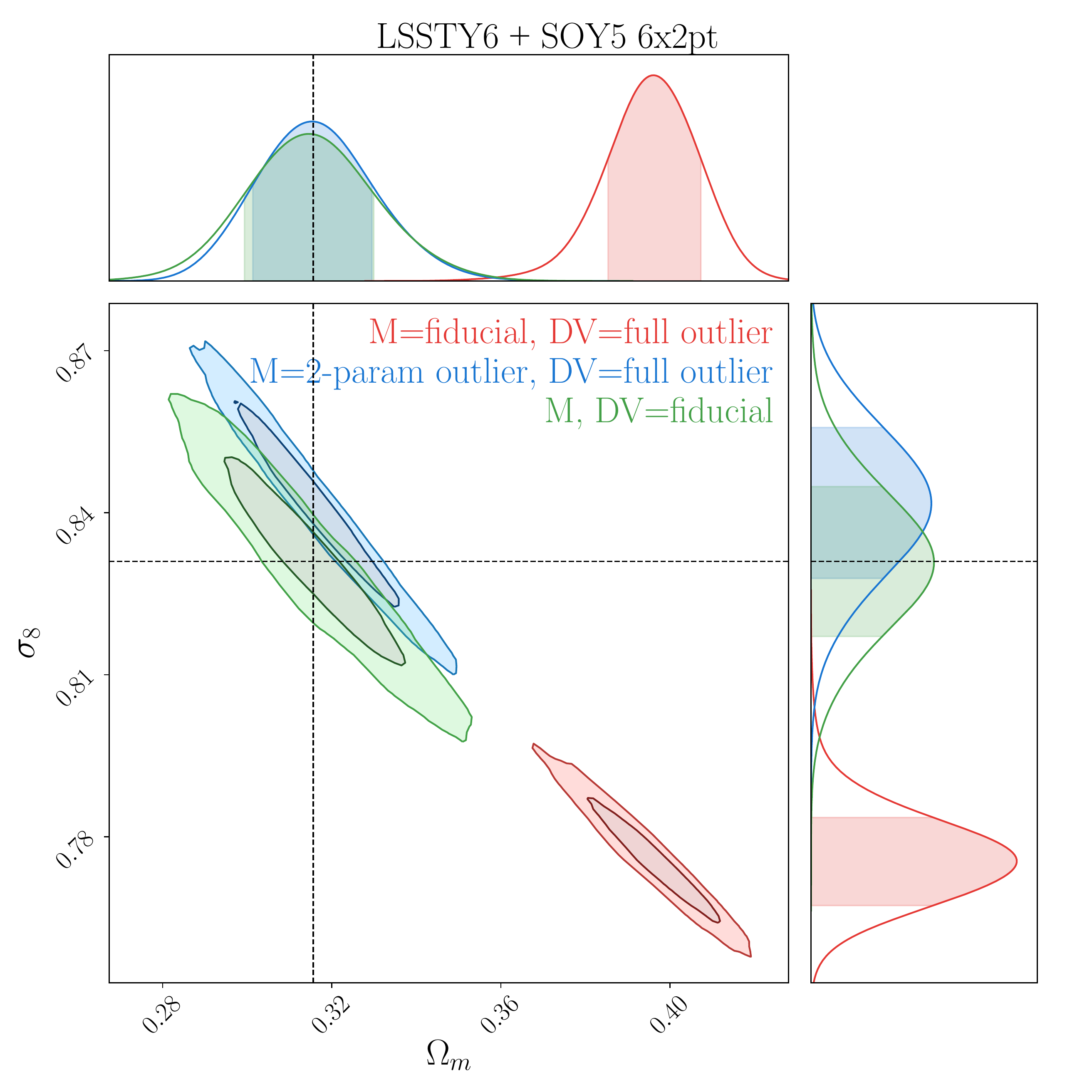}
    \includegraphics[width=0.45\textwidth]{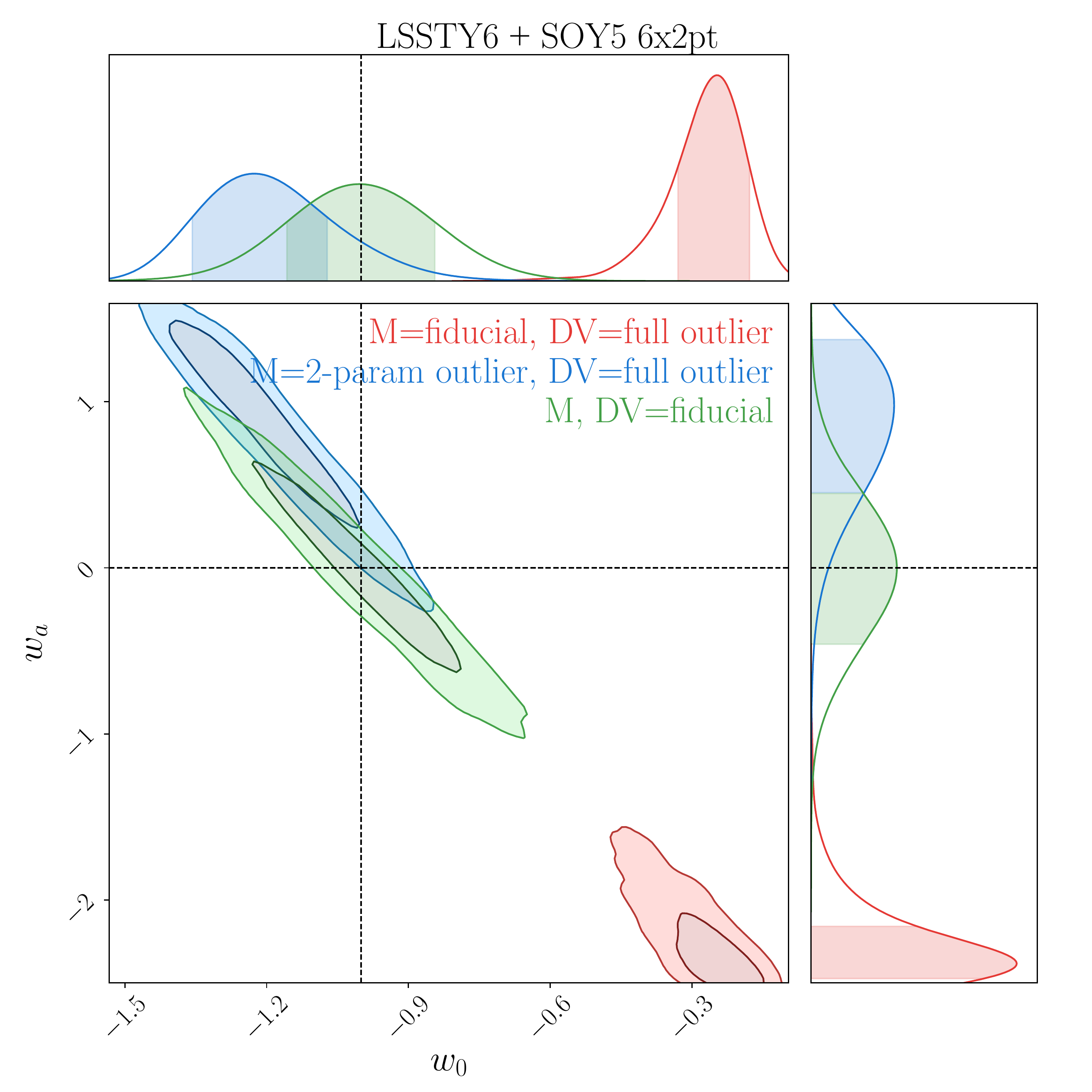}
    \caption{LSST Y6 + SO Y5 6\x2pt constraints of $(\Omega_m,\sigma_8)$ (left) and $(w_p,w_a)$ (right) assuming $w_0-w_a$CDM model. The red and blue contours represent the constraints from fitting the ``full-outlier'' contaminated data vector (DV) with the fiducial model and the 2-parameter outlier model, respectively. For comparison, the green contour shows the constraints from fitting the fiducial DV with the fiducial model. The outlier patterns from the simulated catalogue show significant impact on the parameter inference when only the Gaussian photo-z model is used. Meanwhile, the 2-parameter outlier model is able to correct most of the parameter shifts produced by the photo-z outliers with negligible degradation of the constraining power. The remaining shifts come from the un-modelled outlier patterns outside the 2 ``islands'' as well as the non-uniformity within the 2 ``islands''.}
    \label{fig:full-outliers}
\end{figure*}

\begin{figure*}
    \centering
    \includegraphics[width=0.45\textwidth]{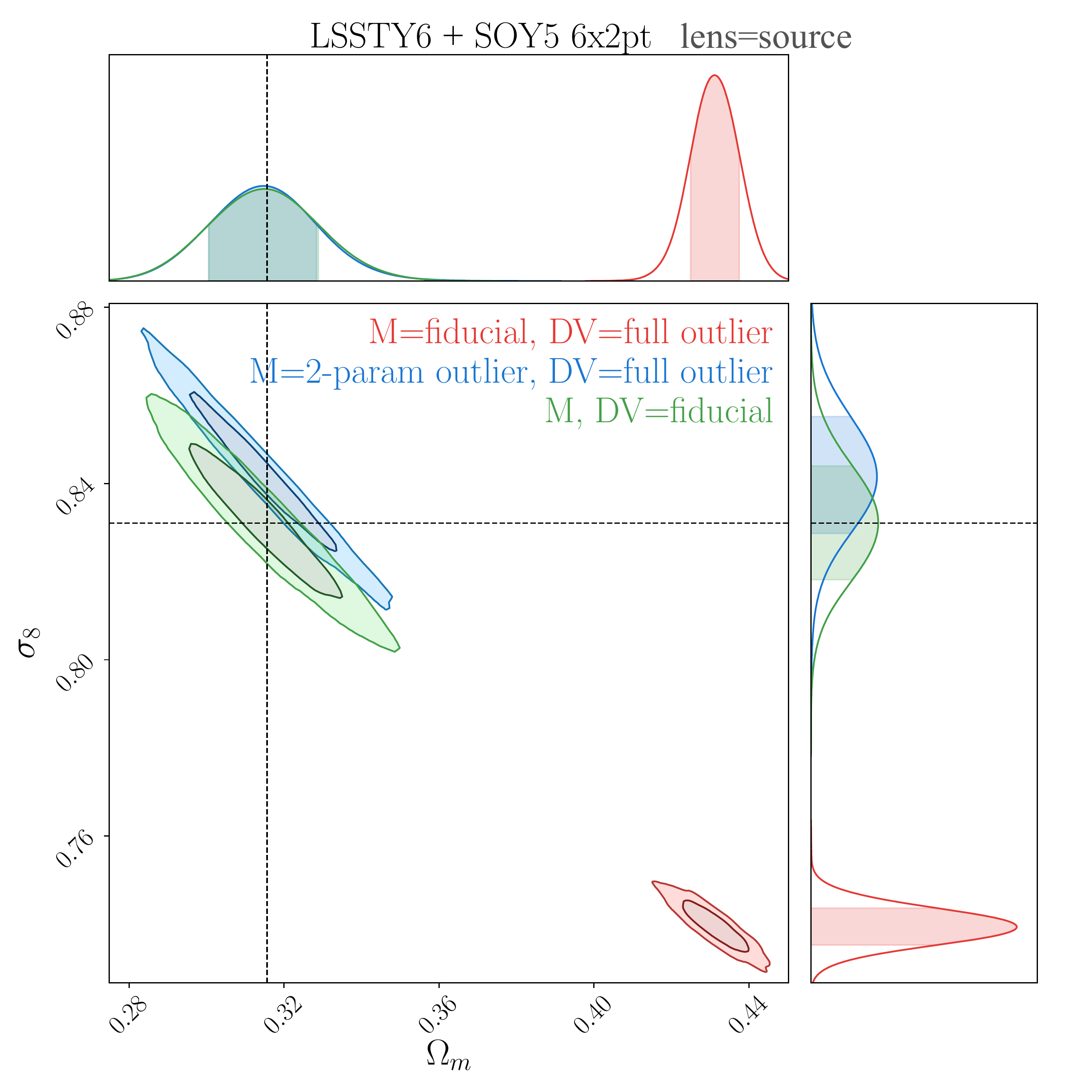}
    \includegraphics[width=0.45\textwidth]{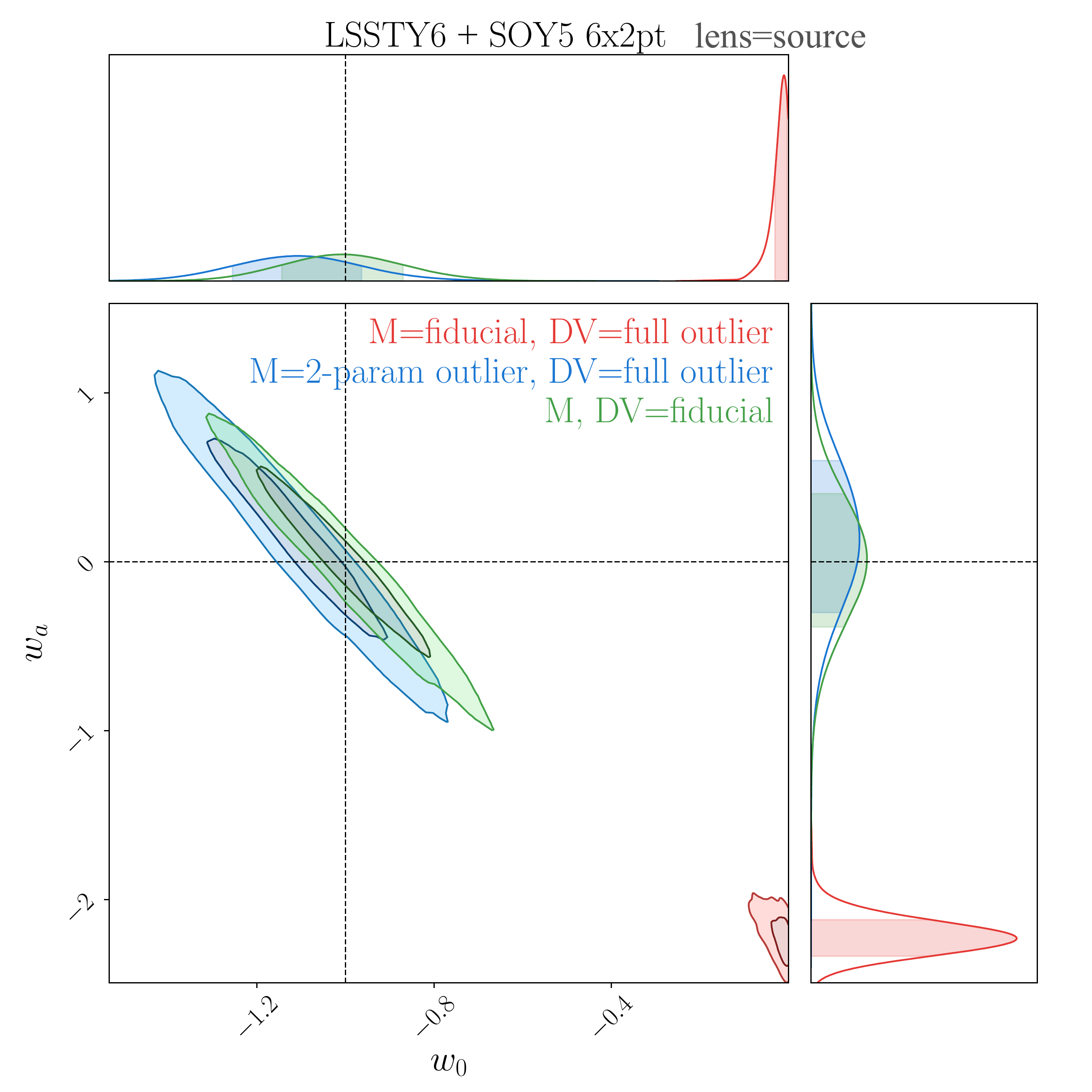}
    \caption{LSST Y6 + SO Y5 6\x2pt constraints of $(\Omega_m,\sigma_8)$ (left) and $(w_p,w_a)$ (right) assuming $w_0-w_a$CDM model and the ``lens=source'' sample choice. The red and blue contours represent the constraints from fitting the ``full-outlier'' contaminated DV with the fiducial model and the 2-parameter outlier model, respectively. For comparison, the green contour shows the constraints from fitting the fiducial DV with the fiducial model. Similar to the SRD case, the outlier patterns from the simulated catalogue show significant impact on the parameter inference when only the Gaussian photo-z model is used. Meanwhile, the 2-parameter outlier model is able to correct most of the parameter shifts produced by the photo-z outliers with negligible degradation of the constraining power. Similar ``island'' modelling scheme can be applied to account for outlier patterns outside the 2 ``islands'', and the non-uniformity within the 2 ``islands'' can be better modelled by sub-dividing the ``islands''.}
    \label{fig:full-outliers_1sample}
\end{figure*}

\begin{figure*}
    \centering
    \includegraphics[width=0.45\textwidth]{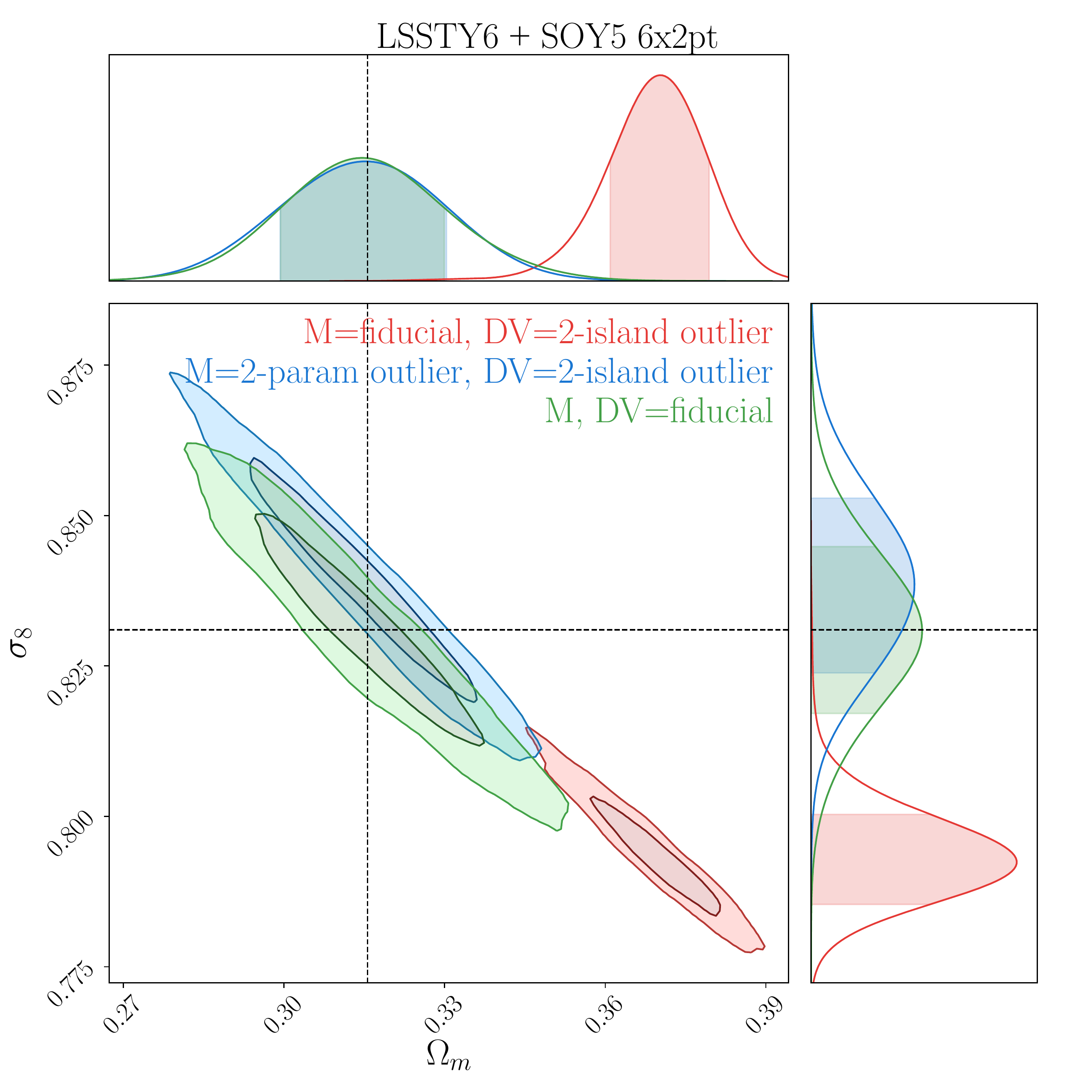}
    \includegraphics[width=0.45\textwidth]{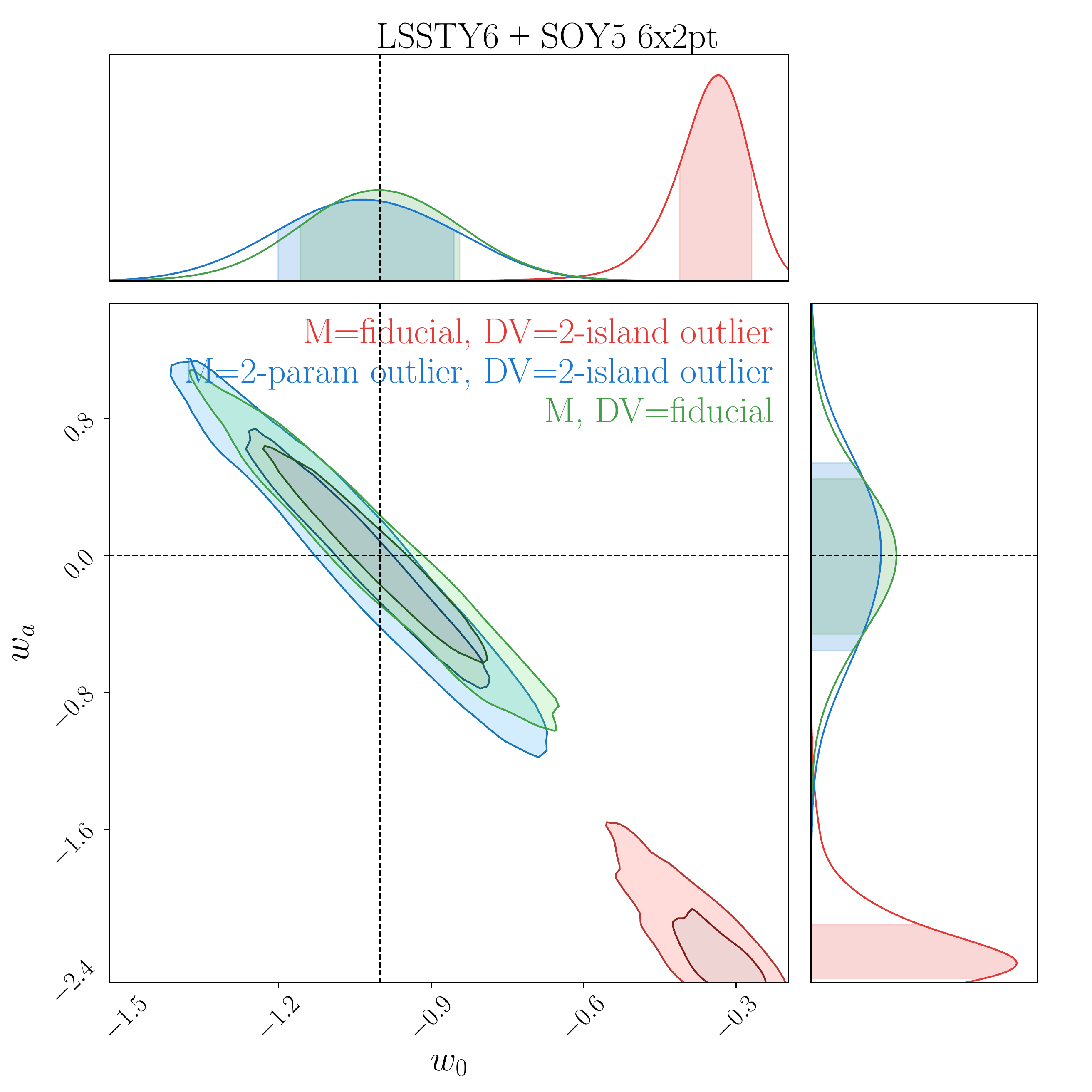}
    \caption{LSST Y6 + SO Y5 6\x2pt constraints of $(\Omega_m,\sigma_8)$ (left) and $(w_p,w_a)$ (right) assuming $w_0-w_a$CDM model. The red and blue contours represent the constraints from fitting the ``2-island'' outlier contaminated DV with the fiducial model and the 2-parameter outlier model, respectively. For comparison, the green contour shows the constraints from fitting the fiducial DV with the fiducial model. The ``2-island'' outlier pattern from the simulated catalogue accounts for most of the impact on the parameter inference presented in the ``full-outlier'' result (Figure~\ref{fig:full-outliers}). The 2-parameter outlier model is able to correct most of the parameter shifts in $(w_0,w_a)$ produced by the ``2-island'' photo-z outliers, while still leaves small shifts in $(\Omega_m,\sigma_8)$. This demonstrates that the assumption (2) that the true redshift distribution given an outlier galaxy’s photometric redshift is uniform within the 2 ``islands'' is a very good approximation, but may still be insufficient for the LSST precision. Further dividing the 2 ``islands'' into sub-islands and applying the same parameterization scheme would be needed to improve the accuracy.}
    \label{fig:outliers-2island}
\end{figure*}

We then follow the steps below to compute $n_{X,\rm out}^i(z)$:
\begin{enumerate}
    \item We obtain the binned photometric redshift distributions $n_X^i(z_{\rm ph})$ as described in Section~\ref{ssec:ana-choices}.
    \item For each bin, we apply the outlier probability matrix $p_{\rm out}(z|z_{\rm ph})$, and obtain the outlier-modified binned true redshift distributions $\tilde{n}_X^i(z)$.
    \item For the fraction of galaxies within the black lines along the diagonal in Figure \ref{fig:z-zph}, we again convolve their distribution with a Gaussian error as described in Eq.~(\ref{eq:binned-nz-true}).
\end{enumerate}
The resulting binned true redshift distribution $n_{X,\rm out}^i(z)$ therefore has an outlier component and a Gaussian component.

We then compute a simulated, outlier infested 6\x2pt data vector at the fiducial cosmological and systematic model (but with the Gaussian + outlier photo-z model).
We analyse the outlier data vectors with the fiducial pipeline (without MG) as presented in Section~\ref{sec:analysis}. The red contours in Figures~\ref{fig:full-outliers} and \ref{fig:full-outliers_1sample} show that the outliers lead to strong biases in cosmological parameters (we consider the $\Omega_m$-$\sigma_8$ plane and the $w_0$-$w_a$ plane) for both the SRD and ``lens=source'' cases, and must be mitigated in future analyses.

\subsection{Outlier mitigation strategy: The island model}\label{ssec:outlier_model}
\cite{2020JCAP...12..001S} present a model describing the redshift mismatch with a mixing matrix $c_{ij}$, the fraction of galaxies in the true redshift bin $i$ that are placed in the photometric redshift bin $j$. In our analysis, this model would require 90 independent parameters for the 10 lens redshift bins. Such a flexible model would more than double the number of free parameters in the analysis, making the inference more challenging, although the additional 90 parameters for the 10 source redshift bins can be avoided if the lens and source samples are chosen to be identical.

We develop an ``island model'' to more efficiently characterise the catastrophic photo-z outliers. In our simulated galaxy catalogue, we have already identified 2 major outlier ``islands'', a region where $z<0.5$ and $z_{\rm ph}>2.0$, and a region where $z_{\rm ph}<0.5$ and $z>2.0$ (marked by the 2 green rectangles in Figure~\ref{fig:z-zph}). 

We note already here that the island model is modular and can be made as complex/flexible as needed. That is additional islands can be added and the distribution within each island can be modified depending on residual biases found in simulated analyses. We further note the need for realistic photo-z simulations, informed by data, that feed information into the development of an optimal island model for LSST. ``Optimal'' here means to retain as much cosmological constraining power as possible, while accounting for biases. 

For now our ``island model'' mitigation strategy assumes that
\begin{enumerate}
    \item outliers are fully described by these 2 islands, and
    \item the true redshift distribution given an outlier galaxy's photometric redshift is uniform, \ie, $p(z|z_{\rm ph})$ is uniform within the outlier island regions.
\end{enumerate}
Since we only consider these 2 islands as outliers, only 2 nuisance parameters are required:
\begin{itemize}
    \item $f_{\rm lowz}$: the fraction of galaxies with $0<z_{\rm ph}<0.5$ that are mis-identified as $2.0<z<3.5$,
    \item $f_{\rm highz}$: the fraction of galaxies with $2.0<z_{\rm ph}<3.5$ that are mis-identified as $0<z<0.5$.
\end{itemize}
With the 2 parameters and the 2 assumptions, we can compute $\tilde{n}_X^i(z)$ and $n_{X,\rm out}^i(z)$ following the same steps as in Section~\ref{ssec:impact}. For the ``full-outlier'' catalogue, we find $f_{\rm lowz}=7.8\%$ and $f_{\rm highz}=13.8\%$ as best fit parameters.

In our analysis, we include $f_{\rm lowz}$ and $f_{\rm highz}$ as additional photo-z parameters to be marginalised over, with prior distributions $[0,20\%]$ and $[0,50\%]$, respectively. We assume that the same photo-z algorithm is used for both the lens and source galaxy samples, so we do not need to introduce an independent set of the free parameters for each sample. Note that for the ``lens=source'' case the assumption is immediately true.

\subsection{Mitigation Results and Tests}\label{ssec:mitigation_tests}
The blue contours in Figures \ref{fig:full-outliers} and \ref{fig:full-outliers_1sample} show that most of the parameter shifts due to the catastrophic photo-z outliers are corrected by the ``2-island model'' for both the SRD and the ``lens=source'' cases. 

The remaining parameter shifts are caused by the inaccuracy of either or both of the 2 assumptions in our ``island model''. To test which assumption is responsible, we need to isolate the impact of the 2 outlier ``islands''. We generate a ``2-island'' outlier catalogue, where only the 2 outlier ``islands'' are considered outliers and all the other galaxies are set to have zero photo-z error and don't enter the probability matrix. Following the steps in Section~\ref{ssec:impact}, we compute the $n_{X,\rm out}^i(z)$ based on this version of outliers. We then compute the outlier-impacted data vector and analyse it with both the fiducial model (Gaussian photo-z only) and the ``2-island model''. The results in Figure~\ref{fig:outliers-2island} show that, if the outliers are solely in the ``2 islands'', then the 2-parameter ``island model'' has been sufficient in recovering the major cosmological parameters $\Omega_m, \sigma_8, w_0$, and $w_a$ to within 2$\sigma$ agreement (Figure \ref{fig:outliers-2island}).

The remaining difference between green and blue contours in Figure~\ref{fig:outliers-2island}, especially in the $(\Omega_m, \sigma_8)$ plane, is a result of the inaccuracy of the assumption (ii), \ie, the uniformity of $p(z|z_{\rm ph})$ within ``islands'', and can be further reduced by defining finer ``islands''.

The difference between the blue contours in Figure~\ref{fig:full-outliers} and in Figure~\ref{fig:outliers-2island} indicates that the assumption (i) needs to be improved, \ie, in addition to the 2 ``islands'' and the Gaussian errors around the diagonal, other outlier structures (\eg, noticeably the ``horns'') in the $z-z_{\rm ph}$ plot (Figure~\ref{fig:z-zph}) need to be accounted for. As a strategy, we suggest applying the same steps to account for them, \ie, for each outlier region, assuming the galaxies at $z_{\rm ph}\in [a_1,a_2]$ have a uniform probability to be mis-identified as $z$ within $[b_1,b_2]$ and introducing an additional ``fraction'' parameter to describe it. The realistic outlier structures depend on the survey characteristics as well as the photo-z estimation algorithm \citep[see \eg,][for a review]{2020MNRAS.499.1587S}. However, we expect that using a few ``fraction'' parameters to approximately describe the major outlier ``islands'' is sufficient to undo the cosmological parameter shifts due to the outliers.

In summary, most of the parameter shifts due to the catastrophic photo-z outliers can be corrected by a simple 2-parameter outlier ``island model'' without degrading the constraining power. The model can be extended to capture more complex outlier structures and improve the accuracy as needed. The exact choice of ``island'' definition and parameterization, as well as the priors on the parameters, should be informed by realistic LSST photo-z simulations. Residual biases must be tested through simulated likelihood analyses. We leave the optimisation of our photo-z outlier modelling strategy for future studies.

\section{Discussion and Conclusion}\label{sec:conclu}

The overlap of large galaxy photometric surveys and CMB experiments in the near future will allow combinations of these datasets to increase the constraining power on cosmological physics. In this paper, we study the joint analysis of LSST and SO, in particular the galaxy position field and lensing shear field from LSST and the CMB lensing field from SO. Our simulated analysis are based on actual MCMC chains, non-Gaussian covariances, extensive systematics modelling of observational (shear calibration and photo-z uncertainties) and astrophysical (galaxy bias, intrinsic alignment, baryonic physics) systematics, and state-of-the-art photo-z simulations for LSST. 

We have covered three main topics in these simulated analyses. First, we compare changes in constraining power when adding SO CMB lensing information to LSST 3\x2pt analyses for a Year 1 and a Year 6 analysis. LSST+SO 6\x2pt presents significant advantages over LSST-only 3\x2pt in constraining dark energy parameters $w_0,w_a$ and MG parameters $\mu_0,\Sigma_0$ (Figure~\ref{fig:3-6x2pt}). In Y1, the dark energy FoM improves by 53\% from 15 (LSST-only 3\x2pt) to 23 (6\x2pt). While in Y6, FoM improves by 92\% from 36 to 69. The MG FoM increases by 72\% in Y1 and 106\% in Y6.

In this context it is important to note that LSST Y1 + SO Y1 6\x2pt is already very promising and demanding as a dataset, measuring parameter $s_8=\sigma_8(\Omega_m/0.3)^{0.35}$ to 0.70\% precision. With increasing depth and survey area, Y6 6\x2pt will see a much tighter constraints on the major cosmological parameters (Figure~\ref{fig:y1-6}), with $s_8$ measured to 0.47\%. The dark energy FoM will triple from 23 to 69.

Our second topic, termed ``lens=source'', is the idea to use the source or weak lensing galaxy sample as the lens or clustering sample. An obvious motivation for this choice is to reduce the dimension of the nuisance parameter space by utilising the identical parameterization of redshift errors for clustering and lensing. We find that ``lens=source'' improves the total SNRs of the measurement by $\sim 30-40\%$ (Table~\ref{tab:snr}). The choice also provides better photo-z calibration (with shift parameters improved by 21-33\% and Gaussian dispersion parameter by a factor of $\sim$7, see Figure~\ref{fig:y6-6x2-1sample-zs}), while mildly improves the cosmological constraints. We note that the higher SNRs will likely benefit the constraints of more complicated physics models beyond time-dependent dark energy.

We stress that some dark energy and MG parameters are measured at better precision with LSST+SO 6\x2pt in Y1 than LSST-only 3\x2pt in Y6. The dark energy equation-of-state at redshift 0.5 is measured to 6.9\% (SRD case) and 5.6\% (``lens=source'') precision in Y1 6\x2pt, while to 6.2\% precision in Y6 3\x2pt. The combined MG parameter $\mu_0+4\Sigma_0$ is measured with a 1$\sigma$ uncertainty 0.23 in Y1 6\x2pt, better than the Y6 3\x2pt uncertainty 0.31 by 26\%. The large constraining power of these joint dataset analyses indicates the near-term need to develop high-precision modelling capabilities and robust systematics mitigation strategies.

Our third and perhaps most interesting topic is exploring catastrophic photo-z outliers, which present a severe obstacle for parameter inference in the LSST-era. We use LSST photo-z mock catalogues to develop a simple but efficient mitigation scheme, the ``island model'', and show that it can correct most of the parameter biases induced by outliers at Y6's precision level. The approach identifies the most important features, in this case the major photo-z outlier ``islands'', from simulations and characterises them with one nuisance parameter per island. In our case we have used two islands and consequently only two additional nuisance parameters to almost fully mitigate the photo-z outlier effects. The outlier ``island model'' can be easily extended to accommodate more complex photo-z scenarios.

Our findings provide strong motivations for the community to prepare for the joint analysis of LSST and SO in the very near future (Y1 joint analysis has significant constraining power). We emphasise that studies of optimal galaxy samples and optimal photo-z outlier mitigation strategies are critically important to extract robust and unbiased cosmological information.

We stress that the constraints presented in this analysis depend on the analysis choices, including the model, parameter priors, survey characteristics, scale range and cuts, etc. Adopting more complex systematic models, \eg, including higher-order perturbation terms for galaxy bias and IA, opens up more degrees of freedom, that can change the relative cosmological constraining power of the probes considered. For example, it can weaken the information from the galaxy survey and thereby making the CMB lensing probes more important. However, we caution against choosing complex models from the start since new degrees of freedom may interplay with each other and projection effects will likely bias the results if the priors are too wide to reject unphysical scenarios. In this paper we adopt systematic models that are simple but flexible enough to describe the systematics identified in current data sets. For future data sets  iterative studies that vary analysis choices are needed to optimise the science return.

The opportunities provided by the combined dataset of galaxy surveys and CMB experiments go beyond the 6\x2pt analysis. Larger multi-probe analyses of CMB\x LSS are possible, \eg, by including the thermal and kinematic SZ (tSZ \& kSZ) effect signals and the cluster counts. Such combinations will allow us to extract cosmological and astrophysical information from even smaller scales, compared to what any individual experiment or probe can achieve.

\section*{Acknowledgements}
We thank Melissa Graham for providing the simulated galaxy photo-z catalogue used in her LSST photo-z studies. We also thank Martin White and an anonymous referee for useful comments on the manuscript.
XF, TE, and HH are supported by the Department of Energy grant DE-SC0020215 and by NASA ROSES ATP 16-ATP16-0084 grant. XF is also supported by the Berkeley Center for Cosmological Physics.
E.S. is supported by the Chamberlain fellowship at Lawrence Berkeley National Laboratory. EK is supported in part by the Department of Energy grant DE-SC0020247, the David \& Lucile
Packard Foundation and the Alfred P. Sloan Foundation.
SF is supported by the Physics Division of Lawrence Berkeley National Laboratory. 
Calculations in this paper use High Performance Computing (HPC) resources supported by the University of Arizona TRIF, UITS, and RDI and maintained by the UA Research Technologies department.

\section*{Data Availability}
The data underlying this paper will be shared on reasonable request
to the corresponding author.

\bibliographystyle{mnras}
\bibliography{references.bib}
\label{lastpage}
\end{document}

%% file: journaldef.tex
\def\aj{AJ}%
\def\apj{ApJ}%
\def\aap{A\&A}%
\def\mnras{MNRAS}%
\def\prd{Phys.~Rev.~D}%
\def\prl{Phys.~Rev.~Lett.}%
\def\pasp{PASP}%
\def\pasj{PASJ}%
\def\jcap{JCAP}%
\def\physrep{Phys.~Rep.}%